\documentclass[usenatbib]{mn2e}
\usepackage{epsfig,times,natbib,aas_macros,rotating}

\begin{document}
\title[Cen A radio-lobe shock]{High-energy particle acceleration at
the radio-lobe shock of Centaurus A} \author[J. H. Croston et al.] {J.
H. Croston$^{1}$\thanks{Email: J.H.Croston@herts.ac.uk},
R.P.~Kraft$^{2}$, M.J.~Hardcastle$^{1}$, M.~Birkinshaw$^{3,2}$,
D.M.~Worrall$^{3,2}$, \newauthor P.E.J.~Nulsen$^{2}$,
R.F.~Penna$^{2}$, G.R.~Sivakoff$^{8,9}$, A.~Jord\'{a}n$^{10,2}$,
N.J.~Brassington$^{2}$, D.A.~Evans$^{4,2}$, \newauthor
W.R.~Forman$^{2}$, M. Gilfanov$^{11,13}$,
J.L.~Goodger$^{1}$, W.E.~Harris$^{5}$, C.~Jones$^{2}$, 
A.M.~Juett$^{6}$, \newauthor S.S.~Murray$^{2}$, S.~Raychaudhury$^{7}$,
C.L.~Sarazin$^{8}$, R. Voss$^{11,12}$ and K.A.~Woodley$^{5}$\\$^1$
School of Physics, Astronomy and Mathematics, University of
Hertfordshire, College Lane, Hatfield, Hertfordshire AL10 9AB\\$^{2}$
Harvard-Smithsonian Center for Astrophysics, 60 Garden Street,
Cambridge, MA~02138, USA\\$^3$ H. H. Wills Physics Laboratory,
University of Bristol, Tyndall Avenue, Bristol BS8 1TL\\$^{4}$
Massachusetts Institute of Technology, Kavli Institute for
Astrophysics and Space Research, 77 Massachusetts Avenue, Cambridge,
MA 02139, USA\\$^{5}$ Department of Physics and Astronomy, McMaster
University, Hamilton, ON L8S 4M1, Canada\\$^{6}$ Laboratory for X-Ray
Astrophysics, NASA Goddard Space Flight Center, Greenbelt, MD
20771\\$^{7}$ School of Physics and Astronomy, University of
Birmingham, Edgbaston, Birmingham B15 2TT, UK\\$^{8}$ Department of
Astronomy, University of Virginia, P.O. Box 400325, Charlottesville,
VA, 22904-4325, USA\\$^{9}$ Department of Astronomy, 4055 McPherson
Laboratory, Ohio State University, 140 West 18th Avenue, Columbus,
OH\\$^{10}$ Departamento de Astronom\'{i}a y Astrof\'{i}sica,
Pontificia Universidad Cat\'olica de Chile, Casilla 306, Santiago 22,
Chile\\$^{11}$ Max-Planck-Institut f\"ur extraterrestrische Physik,
Giessenbackstrasse, D-85748, Garching, Germany\\$^{12}$ Excellence
Cluster Universe, Technische Universit\"at M\"unchen, Boltzmannstr. 2,
D-85748, Garching, Germany\\$^{13}$ Space Research Institute, Russian
Academy of Sciences, Profsoyuznaya 84/32, 117997 Moscow, Russia}
\maketitle

\pagerange{\pageref{firstpage}--\pageref{lastpage}}
\pubyear{2008}

\label{firstpage}

\begin{abstract}
We present new results on the shock around the southwest radio lobe of
Centaurus A using data from the {\it Chandra} Very Large Programme
observations (740 ks total observing time). The X-ray spectrum of the
emission around the outer southwestern edge of the lobe is well
described by a single power-law model with Galactic absorption --
thermal models are strongly disfavoured, except in the region closest
to the nucleus. We conclude that a significant fraction of the X-ray
emission around the southwest part of the lobe is synchrotron, not
thermal. We infer that in the region where the shock is strongest and
the ambient gas density lowest, the inflation of the lobe is
accelerating particles to X-ray synchrotron emitting energies, similar
to supernova remnants such as SN1006. This interpretation resolves a
problem of our earlier, purely thermal, interpretation for this
emission, namely that the density compression across the shock was
required to be much larger than the theoretically expected factor of
4. We describe a self-consistent model for the lobe dynamics and shock
properties using the shell of thermal emission to the north of the
lobe to estimate the lobe pressure. Based on this model, we estimate
that the lobe is expanding to the southwest with a velocity of
$\sim$2600 km s$^{-1}$, roughly Mach 8 relative to the ambient medium.
We discuss the spatial variation of spectral index across the shock
region, concluding that our observations constrain $\gamma_{max}$ for
the accelerated particles to be $\sim$10$^8$ at the strongest part of
the shock, consistent with expectations from diffusive shock
acceleration theory. Finally, we consider the implications of these
results for the production of ultra-high energy cosmic rays (UHECRs)
and TeV emission from Centaurus A, concluding that the shock front
region is unlikely to be a significant source of UHECRs, but that TeV
emission from this region is expected at levels comparable to current
limits at TeV energies, for plausible assumed magnetic field strengths.
\end{abstract}

\begin{keywords}
galaxies: active -- galaxies: elliptical and lenticular --
galaxies:individual (Cen A) -- radio continuum: galaxies -- X-rays:
galaxies -- shock waves 
\end{keywords}

\section{Introduction}
\label{intro}
The mechanisms by which radio-loud active galaxies transfer energy to
their surroundings are of considerable interest, due to the likely
r\^{o}le of such energy input in counteracting gas cooling in the
central regions of galaxies, galaxy groups and clusters. Shocks are
particularly important, as they provide a means of increasing the
entropy of the surrounding gas. There are now a number of examples of
shocks associated with radio galaxies, which include weak shock
structures in the intracluster medium (ICM) of some of the nearest
galaxy clusters \citep[e.g.][]{fab03,fab06,for05} generally on scales
larger than the radio lobes, and a small number of examples of strong
shocks associated with galaxy-scale radio lobes in comparatively
isolated environments \citep[e.g.][]{kra03,kra07a,jhc07,jet08}.

The best-studied example of a strong shock associated with a radio
galaxy is the prominent shell of bright X-ray emission surrounding the
south-west inner lobe of Centaurus A (Cen A), detected with {\it
Chandra} and {\it XMM-Newton}, as first reported by \citet{kra03}. As
Cen A (NGC\,5128) is the nearest radio galaxy ($D_{L} = 3.7$
Mpc, see below), it is an ideal laboratory for investigating the
physics of radio-lobe shocks and their effects on the galaxy
interstellar medium (ISM). Kraft et al. (2003) interpreted the
bright X-ray emission as thermal in origin, based on acceptable fits
to a thermal spectrum, and a lack of radio emission at the location of
the shock front (as might be expected if the X-ray emission was
associated with high-energy particle acceleration); however, their
analysis raised questions about the shock hydrodynamics, as the
apparent density contrast between shocked and unshocked material was
significantly higher than the expected value of 4 for a strong shock
(where the adiabatic index is 5/3). In \citet{kra07a}, deeper {\it
Chandra} observations were used to resolve the structure of the shock
front in more detail, and to investigate the shock structure as a
function of position. They developed a more complex two-fluid model
for the shocked material, which accommodated the electrons and ions in
the shocked material being out of equilibrium. However, that work did
not address the issue of the density contrast between pre-shocked and
post-shocked gas.

Here we present a new analysis of the Cen A shock properties, based on
very deep {\it Chandra} observations of the source as part of a Very
Large Programme carried out in {\it Chandra} Cycle 9. A total of $\sim
740$ ks of {\it Chandra} data on Cen A is now available by combining
all of the available {\it Chandra} data. This rich dataset has already
been used to carry out detailed studies of the X-ray binary population
in NGC\,5128 \citep{jor08,siv08}, particle acceleration in the X-ray
jet \citep{mjh07,wor08}, and of the galaxy ISM \citep{kra08}. Our
analysis in this paper of the new deep observations of the Cen A shell
revises some of the conclusions of earlier work by Kraft et al.

In Section~\ref{analysis} we summarize the data reduction and
analysis, in Section~\ref{results} we discuss the results of our new
spatial and spectral analysis of the X-ray shell, and in
Section~\ref{discuss} we compare our results with previous work,
present a new interpretation for the majority of the shell X-ray
emission, and discuss the implications of that interpretation. We
adopt a distance of 3.7 Mpc to Cen A (the average of five distance
estimates to Cen A -- see Section 6 of \citealt{fer07}), giving an
angular scale of 1 arcsec = 17.9 pc.

\begin{figure}
\begin{center}
\epsfig{figure=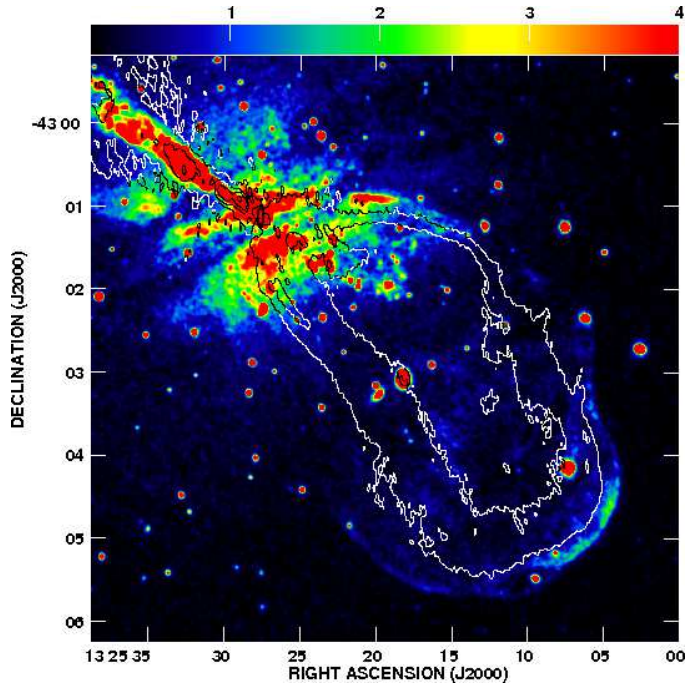,height=9cm}
\caption{The shell of X-ray emission surrounding the SW lobe, from the
merged {\it Chandra} data from all observations to date, in the 0.4 --
2.5 keV energy range. The image has been lightly smoothed with a
Gaussian kernel of FWHM 2.3 arcsec. Radio contours at 1.4 GHz are
overlaid with contour levels $9.0 \times (1, 4, 16, 64)$ mJy
beam$^{-1}$. The colour scale is linear in units of ACIS counts/pix.}
\label{image}
\end{center}
\vskip -15pt
\end{figure}

\section{Data preparation and analysis}
\label{analysis}

Cen A has now been observed a total of 10 times using {\it
  Chandra}'s AXAF CCD Imaging Spectrometer (ACIS) instrument. The
  south-west radio lobe and surrounding X-ray shell are entirely
  within the detector area for 7 of these observations (ObsIDs 0962,
  2978, 3965, 7797, 7800, 8489 and 8490). The effective on-source time
  for the entire shell region is $\sim 507$ ks. The inner parts of the
  shell have a higher total exposure time, but for this analysis we
  chose to include only the datasets in which the entire shell falls
  within the detector field of view. The datasets were all processed
  uniformly using {\sc ciao} version 3.4 and {\sc caldb} version
  3.3.0.1. Further details of the data processing are given in
  \citet{siv08} and \citet{mjh07}. Spectral fitting was carried out
  using {\sc xspec}.

Since the primary beam of the VLA limits the sensitivity of maps of
Cen A at high frequencies and large distances from the pointing
center, we made a new radio map with an effective frequency of 1.5 GHz
from archival VLA datasets. We combined data from observations with
the VLA in its A configuration (AB257: 1983 Oct 30), B configuration
(AC175: 1986 Aug 11), BnC configuration (AC175: 1986 Oct 03) and CnD
configuration (AB587: 1991 Feb 16) to give a map that had the highest
possible resolution while sampling the large-scale structure to the
extent possible with the VLA at this frequency. Calibration and data
reduction were carried out with AIPS in the standard way. The
resulting map has a resolution of $6.1'' \times 1.9''$ (FWHM of major
and minor axes of elliptical Gaussian: the major axis is closely
aligned with the north-south axis) and is dynamic-range limited by
artefacts around the bright core, with an off-source r.m.s. around 1.8
mJy beam$^{-1}$. Images presented in this paper are corrected for
attenuation due to the primary beam of the VLA.
\begin{figure}
\begin{center}
\epsfig{figure=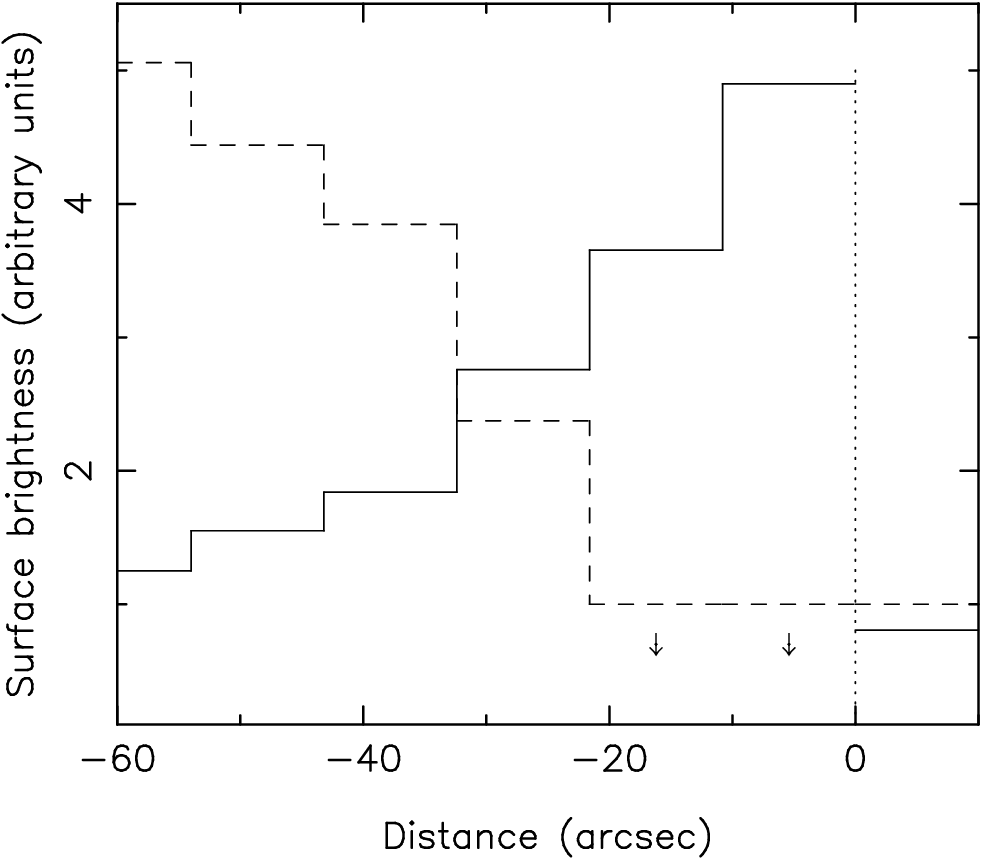,height=7cm}
\caption{X-ray (solid line) and radio (dashed line) surface brightness
  profiles across the shock front (the extraction regions are shown in
  Fig.~\ref{regions}). Arrow symbols indicate positions at which the
  radio profile is based on (3$\sigma$) upper limits. The profiles
  have been renormalised for ease of comparison. The dotted line
  indicates the position of the shock front as determined from the
  X-ray data. The bin width is significantly larger than the size of
  the radio beam, so that all data points are independent.}
\label{profile}
\end{center}
\end{figure}
\section{Results}
\label{results}

\begin{figure*}
\caption{Source and background regions used for the spectral and
  spatial analysis. Top left: Region 1, which includes the brightest
  part of the shell emission; top right: Regions 2 and 3: the region
  of X-ray shell emission exterior to the radio lobe (bottom right of
  the image) and the region of bright emission at the north-east edge
  of the lobe, respectively; bottom left: zoomed-in view of Region 3,
  the north-east edge of the SW shell; bottom right: regions used for
  surface brightness profile in Fig.~\ref{profile}. Chip gaps and out of
  field-of-view regions were masked out of background spectra on an
  observation by observation basis as necessary, and point sources
  were excluded from all spectra.}
\label{regions}
\centerline{\vbox{\hbox{
\epsfig{figure=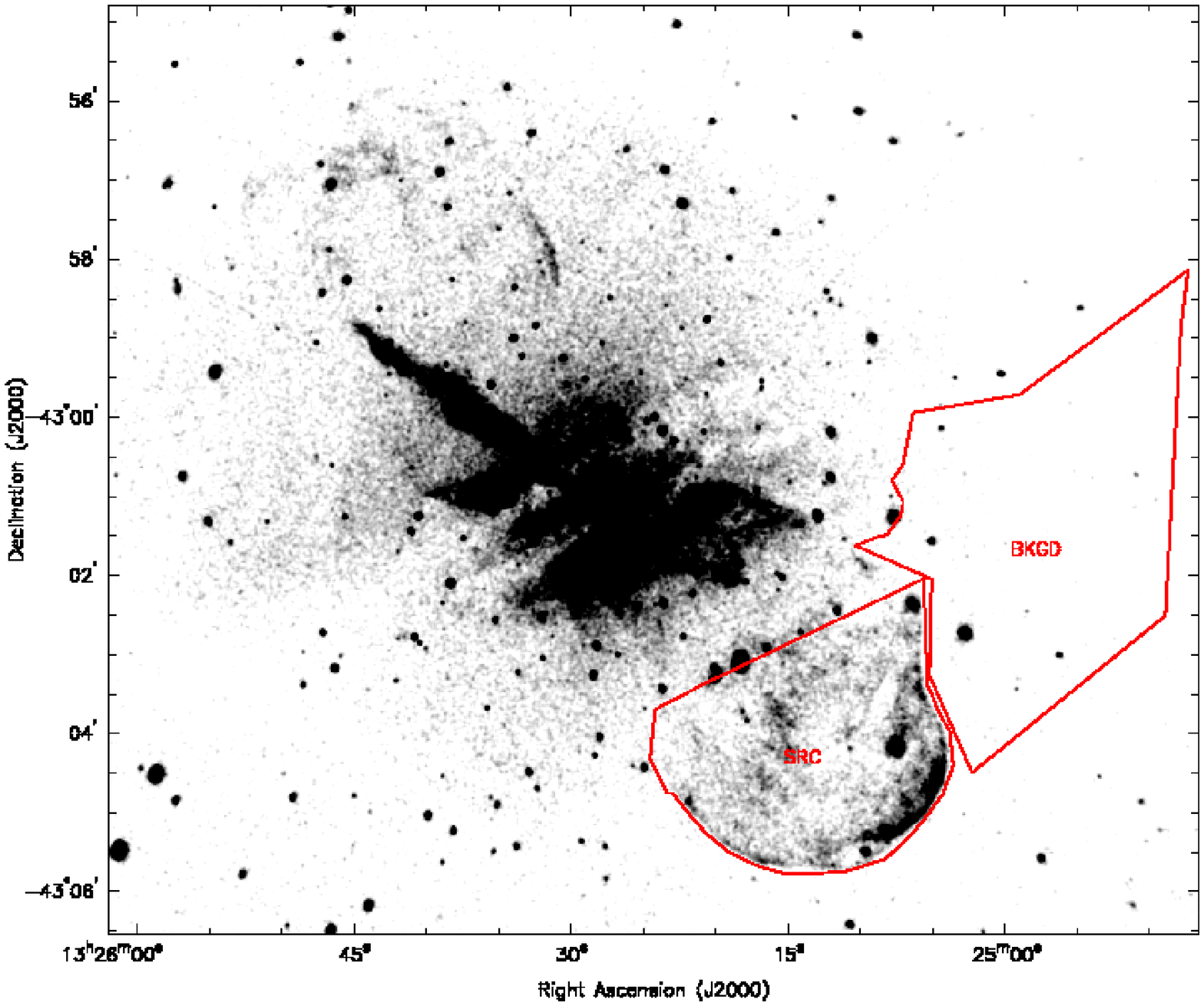,height=6cm}
\epsfig{figure=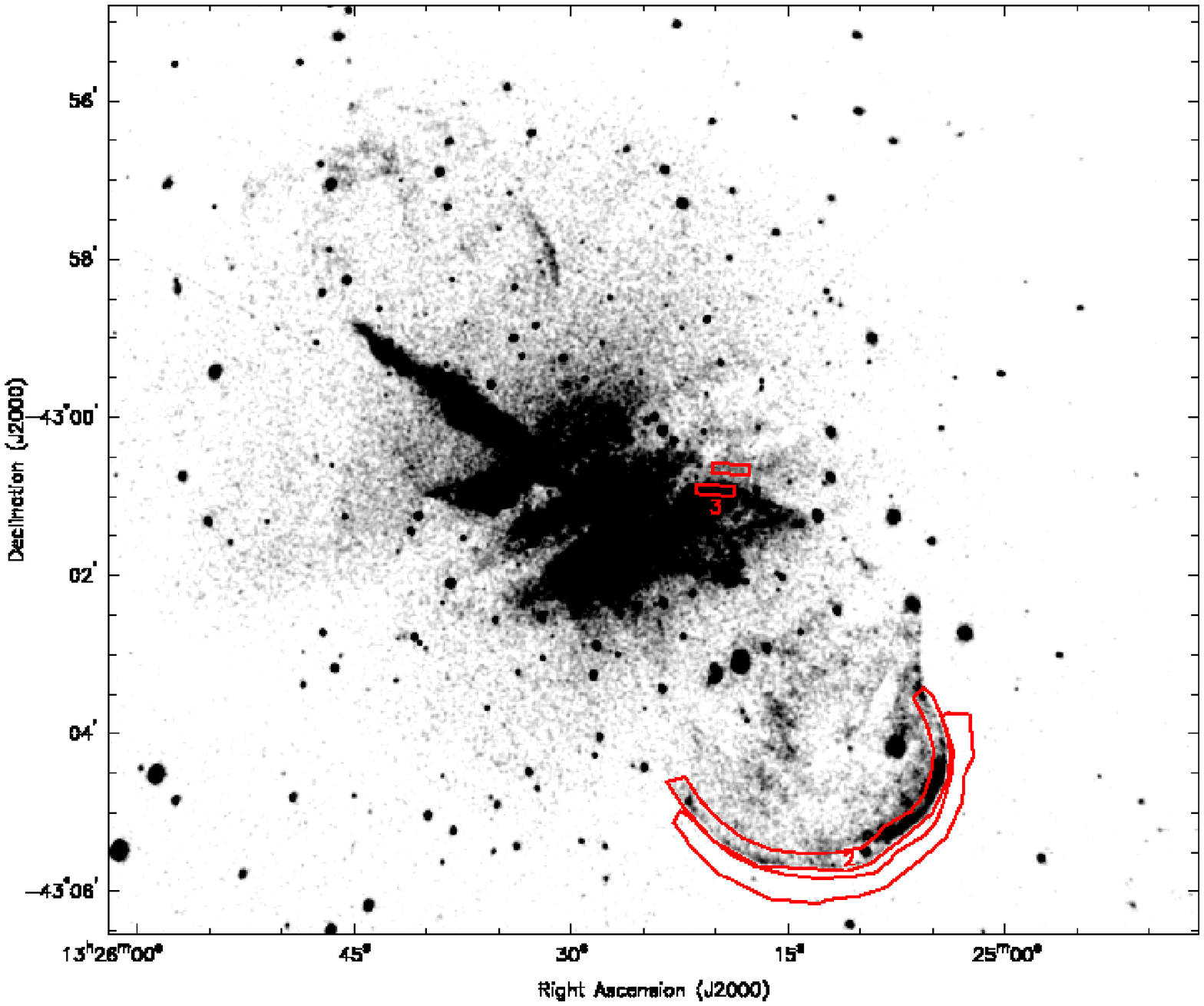,height=6cm}}
\hbox{
\epsfig{figure=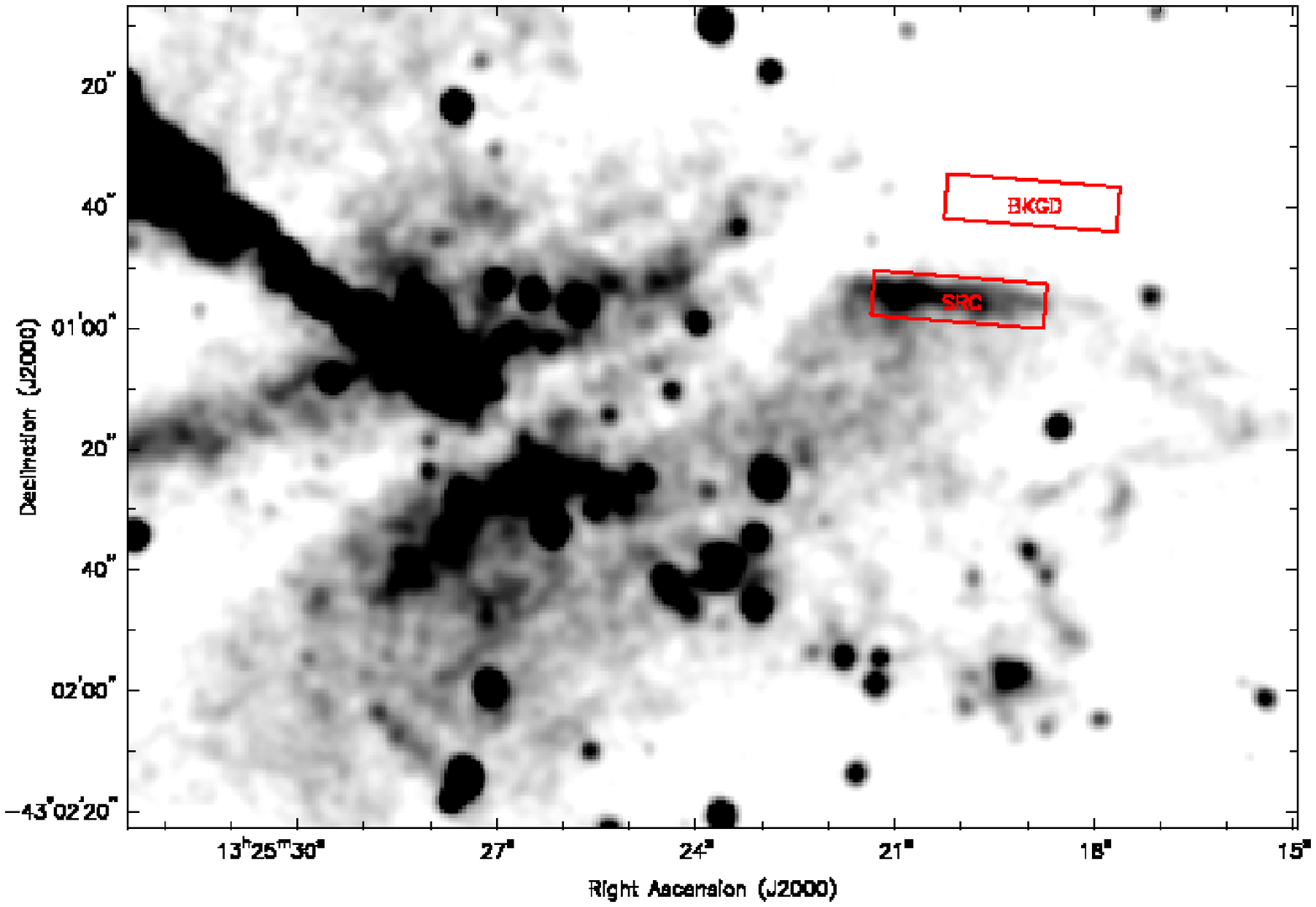,height=5cm}
\epsfig{figure=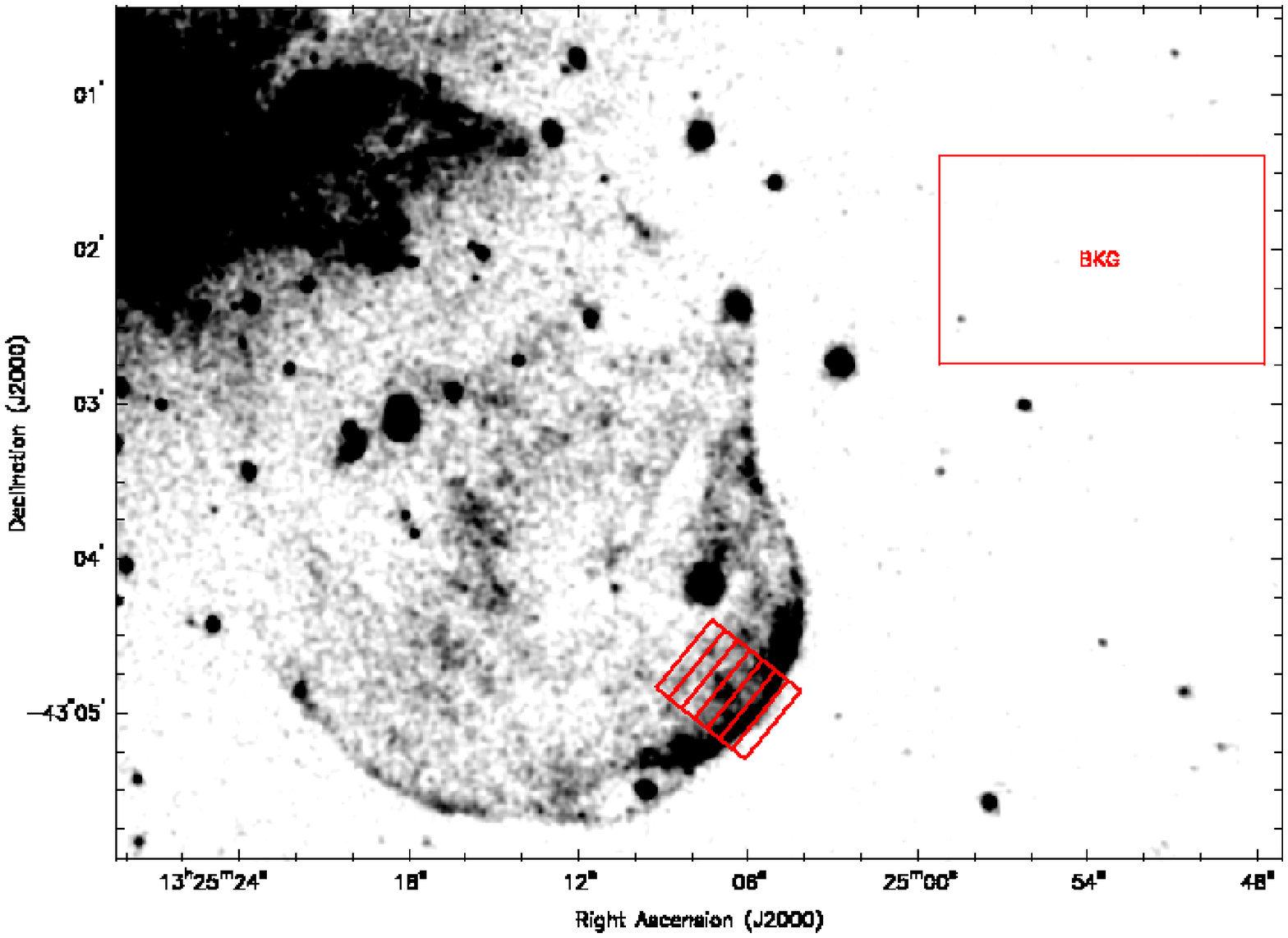,height=5cm}}}}
\end{figure*}

Fig.~\ref{image} shows a Gaussian-smoothed image of the X-ray emission
surrounding the south-west radio lobe with radio contours overlaid. It
can be seen that the shell is sharply edge-brightened. We interpret
the outermost edge of the X-ray emission as the shock front. This lies
$\sim 16$ arcsec from the outer edge of the radio lobe, which is well
defined in our radio images, as demonstrated by the radio and X-ray
surface brightness profiles shown in Fig.~\ref{profile}, extracted for
the regions shown in the bottom right panel of Fig.~\ref{regions}. We
interpret the sharp radio lobe boundary to be the contact
discontinuity. Shell emission is seen throughout the lobe region,
including a bright region on the northern edge of the shell, a
distance of 1.4 arcmin E of the nucleus, which was first discussed by
\citet{kra07a}. The fact that the X-ray emission clearly extends a
significant distance beyond the edge of the radio lobe (well defined
in our high resolution radio maps) and is edge-brightened (see
Fig.~\ref{profile}) strongly supports a model in which the X-ray
emission originates in a shell surrounding the radio lobes, rather
than originating in a fluid contained within them.

We initially carried out spectral analysis for several large regions
to determine the global properties of the shell X-ray emission. The
spectral extraction regions are shown in Fig.~\ref{regions}. Spectra
were extracted from all seven datasets using the {\sc ciao} task {\it
specextract}, which also calculates appropriate response files. The
background regions used are also shown in Fig.~\ref{regions}. For the
background spectra, additional exclusion regions were included on an
observation-by-observation basis to mask any out-of-field-of-view
regions and chip gaps. Point sources were also excluded from all of
the extracted spectra. The spectra were fitted in the 0.4 -- 3.0 keV
energy range to minimize background contamination from the wings of
the PSF from the very bright central AGN \citep[e.g.][]{eva04}, which
dominate above 3.0 keV across the entire field-of-view
\citep[e.g.][]{kra07a}. This choice of energy range also minimizes any
contamination from readout streaks. In the few cases where readout
streaks pass through regions of interest (e.g. ObsID 7800 for Region
3, and ObsIDs 7797 and 8490 for a small number of inner regions
discussed in Section~\ref{spix}) we checked that excluding the datasets
that could be contaminated did not affect the spectral results. For
each region we carried out joint fits to the spectra from all epochs,
and we assumed a fixed hydrogen column density at the Galactic value
of N$_{H} = 8.4 \times 10^{20}$ cm$^{-2}$ \citep{dl90}, except where
otherwise specified. This is reasonable in general, as our spectral
extraction regions are mainly well away from the central region where
varying absorption from gas associated with the dust lane needs to be
taken into account. For each of the regions we initially fitted five
spectral models: an APEC (Astrophysical Plasma Emission Code:
\citealt{smi01}) model with free abundance (Model I), an APEC model
with abundance fixed at $0.15Z_{\sun}$ (Model II), an APEC model with
abundance fixed at $0.5Z_{\sun}$ (Model III), a single power-law model
(Model IV) and a model consisting of an APEC ($Z = 0.5$) plus a
power-law component (Model V). The abundance values for Models II and
III were chosen to sample the range of abundances from the low values
found in free-abundance fits to more typical values for galaxy
atmospheres \citep[e.g.][]{dav06}. Complete results of all the
spectral fits are given in Table~\ref{spectra}. Quoted errors are
$1\sigma$ for one or two interesting parameters except where otherwise
specified.

\subsection{The bright outer shell emission}
\label{outer}
\begin{figure*}
\centerline{\vbox{\hbox{
\epsfig{figure=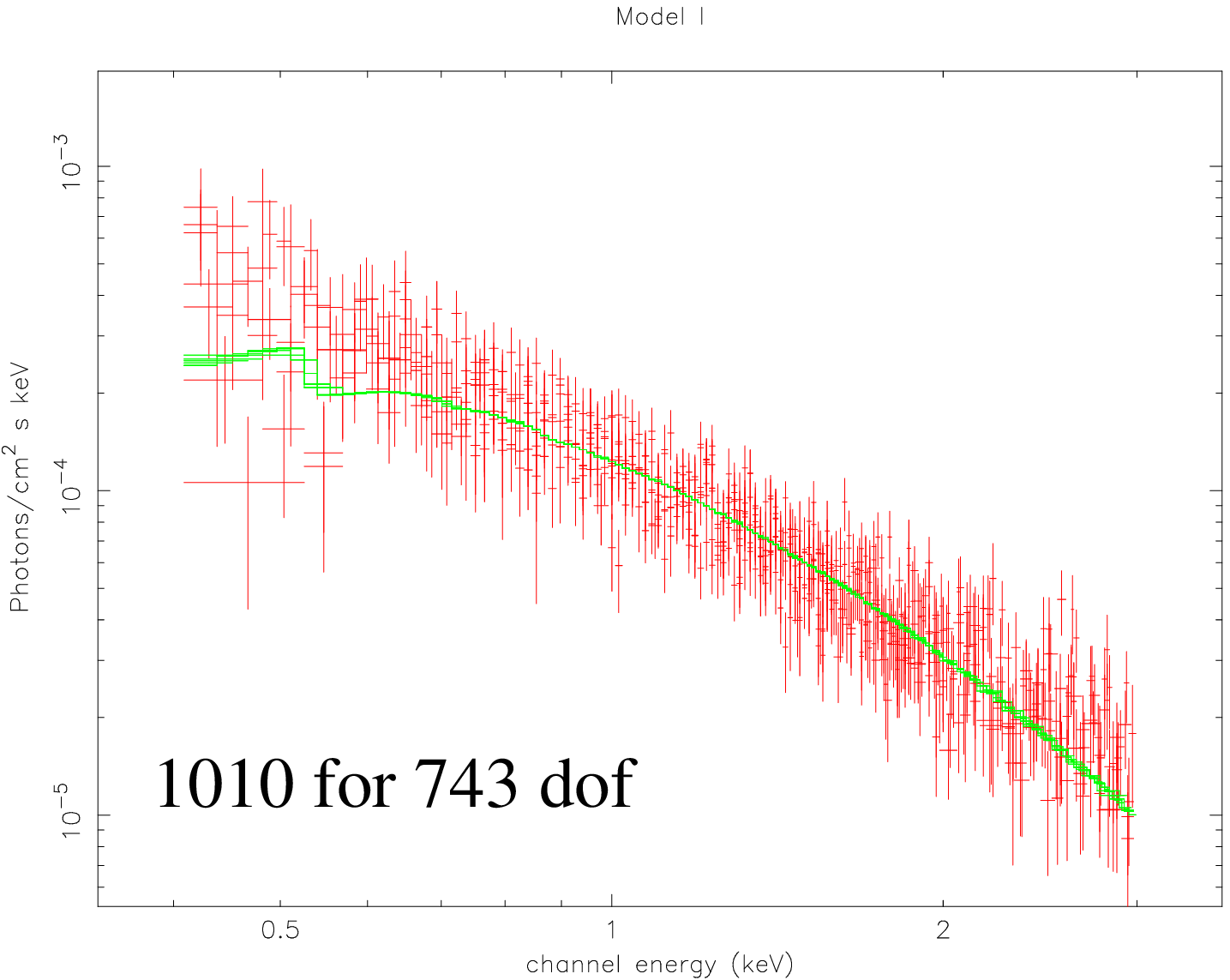,height=4.5cm}
\epsfig{figure=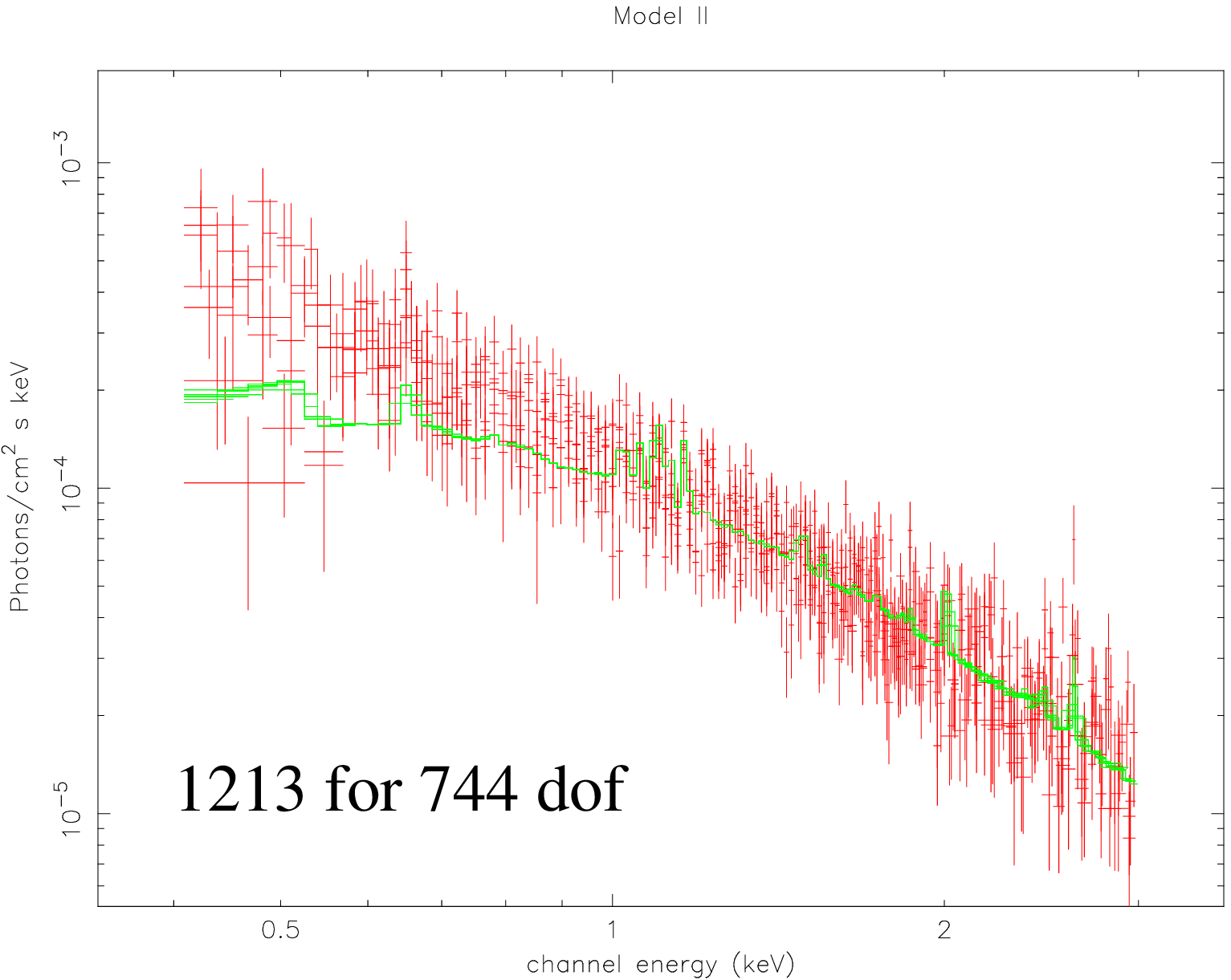,height=4.5cm}
\epsfig{figure=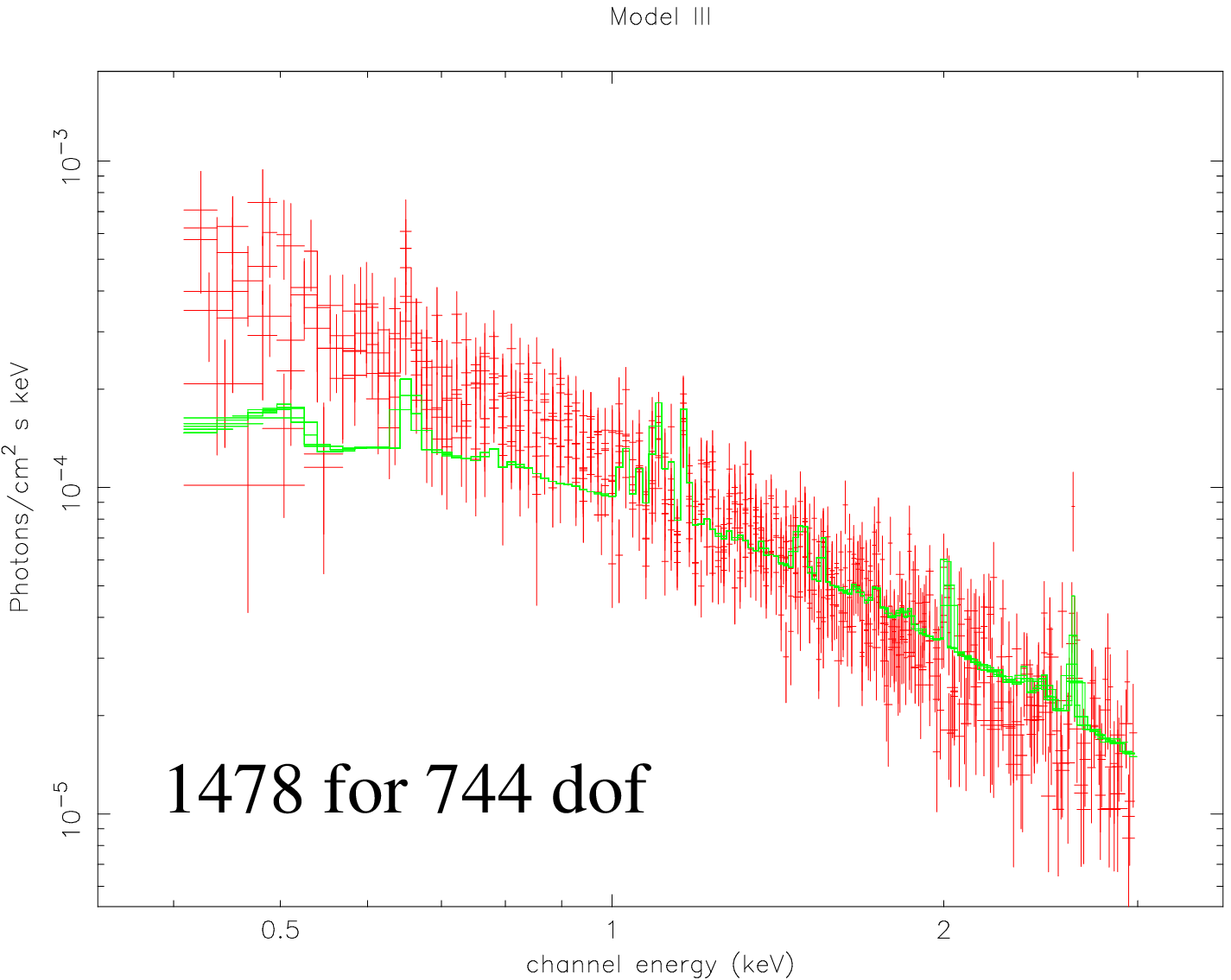,height=4.5cm}}
\hbox{
\hskip 2.5cm
\epsfig{figure=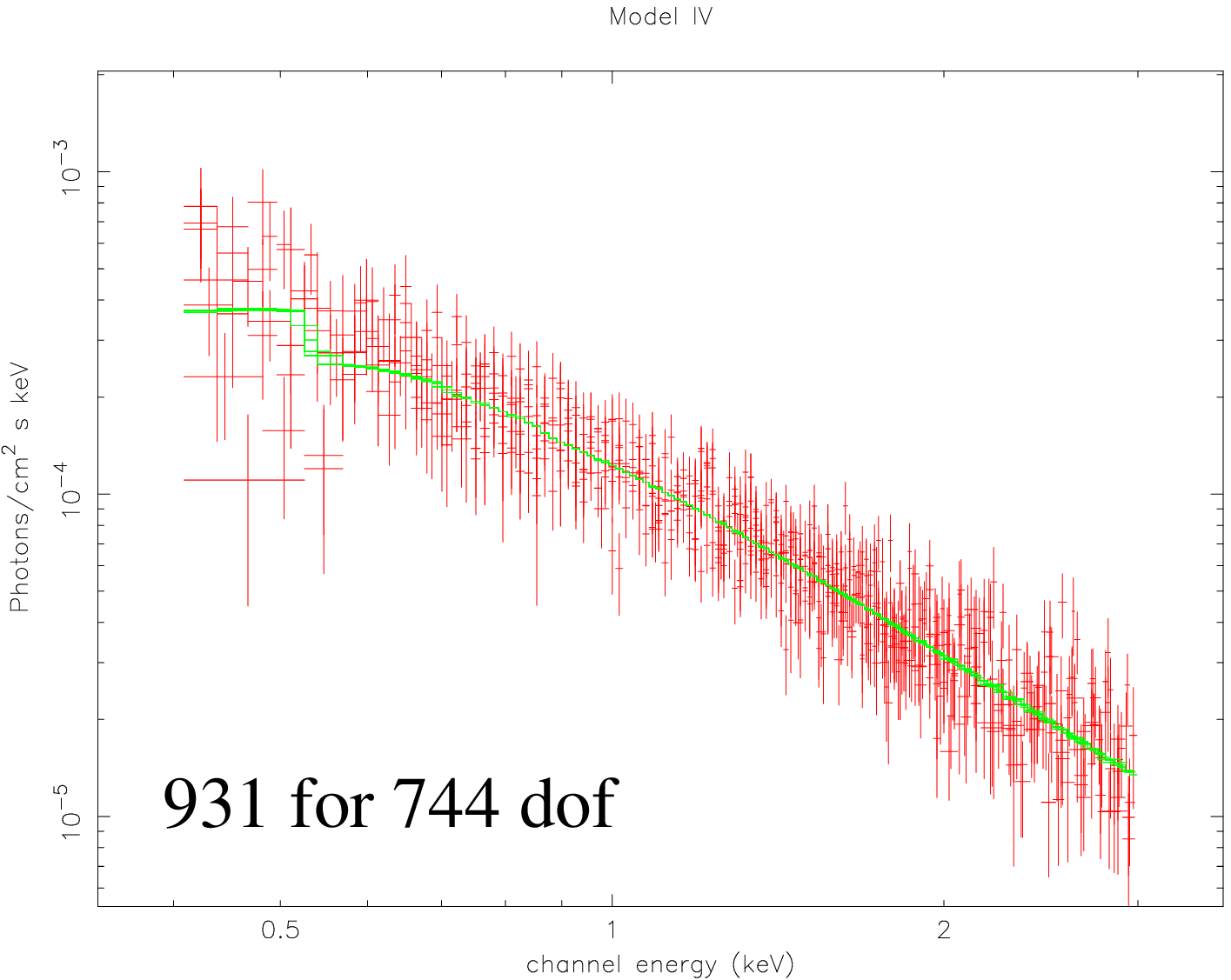,height=4.5cm}
\epsfig{figure=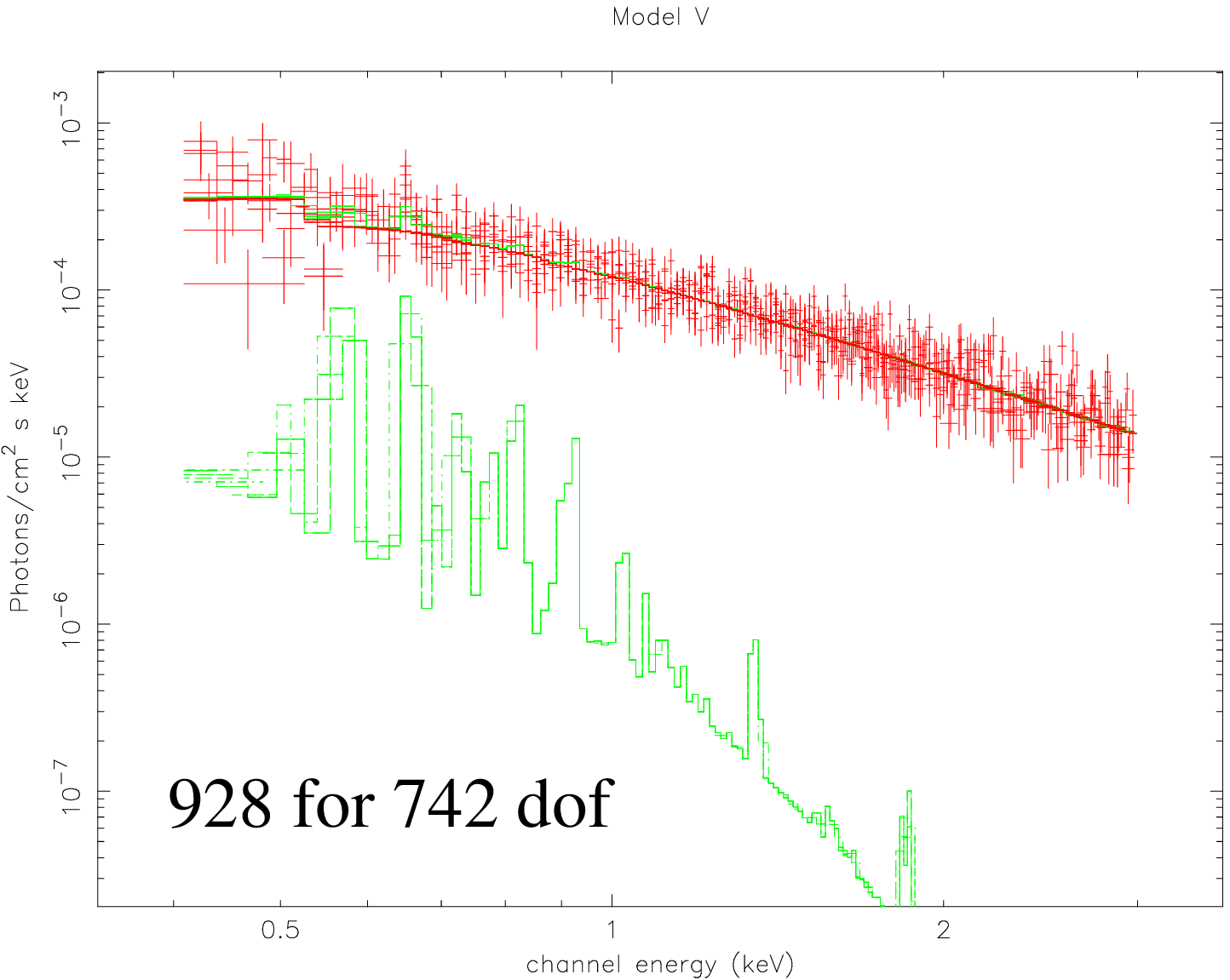,height=4.5cm}}}}
\caption{Spectral fits for Region 1. Top left: Model I (free-abundance
  APEC), middle: Model II (APEC with $Z = 0.15 Z_{\sun}$), right:
  Model III (APEC with $Z = 0.5 Z_{\sun}$), bottom left: Model IV
  (power law), right: Model V (power law + fixed-abundance APEC). The
  data are in red and the fitted models are in green. The y-axis scale
  is different for Model V due to the inclusion of the thermal
  component, which has a low flux.}
\label{mod1}
\end{figure*}

We first considered a region encompassing all of the bright shell
emission (Region 1 in Fig.~\ref{regions}), with point sources
excluded. While the spectral model parameters vary somewhat across
this large region, the general features of the model fits described
below are present in all the fits from smaller regions when Region 1
is subdivided. For this region we find that Models IV and V (single
power law and 2-component power-law plus APEC, respectively) give the
best fits to the spectra. Models I -- III (thermal models) give much
poorer fits. Model I, the free abundance fit, has the lowest
$\chi^{2}$ of all of the thermal models; however, the abundance is
constrained to be extremely low (a $3\sigma$ upper limit on $Z$ of
0.007 $Z_{\sun}$), and the fitting statistic is still significantly poorer
than for the power-law fit (Model IV). For Model V (APEC plus power
law), the inclusion of a thermal component does not significantly
improve the fit when compared to Model IV, and the thermal model
contributes an unabsorbed 0.4 -- 3.0 keV flux of $\sim 1.6 \times
10^{-14}$ erg cm$^{-2}$ s$^{-1}$, whereas the power-law contributes a
flux of $\sim 4.7 \times 10^{-13}$ erg cm$^{-2}$ s$^{-1}$, a factor of
30 times higher. The spectrum for Region 1 is shown in Fig.~\ref{mod1}
with each of the five models overplotted. The fitting statistics for this
region are fairly poor for all five models. This is likely to be due
to radial variation in the spectral properties of the emission,
possibly including variation in the contribution from thermal
emission. We explore this spatial variation in Section~\ref{spix}.

To summarize, we find that a single power law is a better description
of the spectrum than any thermal models. In addition, the fitted
temperature values for thermal models are strongly dependent on the
adopted abundance. The shallower observations analysed by
\citet{kra03,kra07a} meant that thermal models provided an adequate
fit to the data, but with our deeper data, we are now able to rule out
a simple, single-temperature APEC model at high confidence. In
particular, the thermal fits have large residuals at soft energies.
Fitting a two-temperature APEC model can mitigate this problem to some
extent: we find best-fitting temperatures of $0.28\pm0.02$ keV and
$3.90\pm0.26$ keV, assuming 0.5 times solar abundances. This fit has
$\chi^{2}$ of 1065 for 742 d.o.f, which is a significant improvement
on the single temperature models for realistic abundances, but still
considerably poorer than the power law fit.

In order to firmly rule out a thermal origin for the shell emission,
we also considered the limits on the 6.7 keV line from ionized Fe, and
the Fe L complex, for assumed gas temperatures of 1.5 and 3 keV
(spanning the range seen in the thermal spectral fits). We found that
the limits on the 6.7 keV line do not provide strong constraints, but
the predicted fluxes in the Fe L complex at 1.17 keV for an abundance
of 0.15Z$_{\sun}$ are $1.6 \times 10^{-5}$ and $4 \times 10^{-6}$
photons cm$^{-2}$ s$^{-1}$, respectively, whereas the $3\sigma$ upper
limit from our joint fits (using the linewidth and energy of a broad
Gaussian fit to the complex in a faked spectrum of the same
parameters) is $\sim 2 \times 10^{-6}$ photons cm$^{-2}$ s$^{-1}$.
Hence there is no evidence for the line emission expected from a
thermal plasma with realistic temperature and abundance values. We
cannot rule out a model in which turbulent broadening has erased the
lines from the spectrum, but this would require turbulent velocities
at or well above the sound speed of the gas.

We also examined separately the brightest region of shell emission at
the south-west rim, $\sim$ 5.6 arcmin to the SW of the nucleus (Region
2 in Fig.~\ref{regions}), which is external to the radio lobes (see
Fig.~\ref{profile}), to construct a broad-band SED for the shell
without contamination from any possible radio lobe emission (see
below). Details of the spectral fits are in Table~\ref{spectra}. The
overall conclusions of the spectral analysis for this region are very
similar to those for Region 1; this is, the spectrum is much better
fitted by a non-thermal power-law component rather than thermal
emission. This is as expected given that this region contributes a
significant fraction of the flux in Region 1.

We therefore conclude that the shell emission is likely to be
non-thermal, in contrast to the conclusions of earlier work based on
poorer quality spectra.

\subsection{The north-east edge of the shell}
\label{s1}
As shown in Fig.~\ref{image}, there is a small region of brighter
shell emission at the northern edge of the SW shell close the AGN
($\sim$ 1.4 arcmin east of the nucleus: Region 3 in
Fig.~\ref{regions}). This region was shown to be associated with the
shell and examined in detail by \citet{kra07a}, who concluded that the
emission was thermal, and significantly cooler than at the outer, SW
edge of the shell ($0.6 - 0.8$ keV for their S1 and S2 regions for an
assumed abundance of $0.5Z_{\sun}$). Since the line emission is
expected to be stronger at these temperatures, this region can
potentially provide useful constraints on the shell X-ray emission. We
therefore examined spectra for Region 3 (Fig.~\ref{regions}), which is
similar to \citet{kra07a}'s S1 region. Results are listed in
Table~\ref{spectra}.

In contrast to the situation in Region 1 and sub-regions of the outer
shell, we find that a thermal model is strongly favoured in this
region. With a free abundance fit, we obtain a best-fitting
temperature of $0.91\pm0.04$ keV and an abundance of
$0.17^{+0.05}_{-0.04}$ Z$_{\sun}$ with $\chi^{2} = 137$ for 97 d.o.f.
The Fe L complex is very prominent in the spectrum, which cannot be
fitted adequately with a power-law model. The inclusion of a power-law
component to the model, in addition to the APEC, does not improve the
fit. Hence we conclude that the X-ray emission from this region is
unambiguously thermal. The best-fitting abundance is significantly
lower than that expected from the galaxy ISM, which is likely to be
due to the assumption of a single temperature \citep[e.g.][]{buo00}.
As our spectral regions are likely to include multi-temperature gas,
we cannot obtain a reliable estimate of the metallicity from the X-ray
data. We therefore chose to adopt a metallicity of $Z = 0.5 Z_{\sun}$,
more appropriate for a galaxy halo. The single temperature fits also
showed strong residuals at $\sim 1.35$ keV, likely to be due to Mg XI,
and so we chose to include a narrow Gaussian at this energy. The best
fitting temperature for an APEC model with fixed abundance plus 1.35
keV Gaussian is $0.95^{+0.03}_{-0.02}$ keV, with $\chi^{2} = 145$ for
97 d.o.f. We also investigated whether additional Silicon features
are present in the spectrum as would be expected if the gas is
overabundant in alpha elements. We found a significant improvement in
the fit
if a Gaussian is added
at 1.865 keV, corresponding to Si XIII, and a marginal improvement
including a Gaussian at 2.006 keV, corresponding to Si XIV. Using the
VAPEC model with all other elements fixed at an abundance of 0.5 times
solar (which gives a very similar fit), we find a magnesium abundance
of $2.4^{+0.3}_{-0.4}$ times solar. If the silicon abundance is
allowed to vary, an abundance of $1.2^{+0.3}_{-0.2}$ times solar is
obtained. The data are therefore consistent with a model in which the
shocked gas in this region is alpha-enriched. It is also possible that
the unexpectedly strong magnesium and silicon lines are indicative of
a non-equilibrium ionization state.

\begin{table*}
\caption{Model fits to X-ray spectra. All fits are joint fits to
  spectra from all seven observations in the energy band 0.4 -- 3.0
  keV. Galactic absorption of $8.4 \times 10^{20}$ cm$^{-2}$ was
  assumed for all of the fits. For Model V we list the inferred,
  unabsorbed 0.4 -- 3.0 keV fluxes for each model component as a
  measure of the relative importance of thermal and non-thermal
  emission in each region.}
 \label{spectra}
\begin{tabular}{llrrrr}
\hline
Model&&Parameter&Region 1&Region 2&Region 3\\
\hline
Model I:&{\it APEC} (free abundance)&$kT$ (keV)&$1.60^{+0.07}_{-0.06}$&$2.03^{+0.24}_{-0.14}$&$0.91^{+0.04}_{-0.04}$\\
&&$Z$ ($Z_{\sun}$)&$<0.007$&$<0.05$&$0.17^{+0.05}_{-0.04}$\\
&&$\chi^{2}$ (d.o.f.)&1010 (743)&344 (308)&137 (97)\\
Model II:&{\it APEC} (fixed abundance)&$kT$ (keV)&$2.22^{+0.08}_{-0.06}$&$2.59^{+0.13}_{-0.13}$&$0.90^{+0.03}_{-0.02}$\\
&&$Z$ ($Z_{\sun}$)&0.15&0.15&0.15\\
&&$\chi^{2}$ (d.o.f.)&1213 (744)&377 (309)&138 (98)\\
Model III:&{\it APEC} (fixed abundance)&$kT$ (keV)&$3.26^{+0.09}_{-0.09}$&$3.58^{+0.24}_{-0.19}$&$0.97^{+0.01}_{-0.02}$\\
&&$Z$ ($Z_{\sun}$)&0.5&0.5&0.5\\
&&$\chi^{2}$ (d.o.f.)&1478 (744)&437 (309)&177 (98)\\
Model IV:&{\it power law}&$\Gamma$&$2.20^{+0.02}_{-0.02}$&$2.01^{+0.04}_{-0.03}$&$2.54^{+0.05}_{-0.05}$\\
&&$\chi^{2}$ (d.o.f.)&931 (744)&333 (309)&445 (98)\\
Model V:&{\it APEC + power law}&$kT$ (keV)&$0.23^{+0.09}_{-0.08}$&0.93$^{a}$&$0.89^{+0.05}_{-0.06}$\\
&&$Z$ ($Z_{\sun}$)&0.5&0.5&0.5\\
&&$\Gamma$&$2.15^{+0.06}_{-0.05}$&$2.00^{+0.15}_{-0.06}$&$2.13^{+0.30}_{-0.27}$\\
&&$\chi^{2}$ (d.o.f.)&928 (742)&332 (307)&141 (96)\\
&&$F_{apec}$&$(1.4^{+0.8}_{-1.2}) \times 10^{-14}$&$<4.8 \times
10^{-14}$&$(2.0\pm0.2) \times 10^{-14}$\\
&&$F_{PL}$&$(4.7^{+0.1}_{-0.2}) \times
10^{-13}$&$(1.38^{+0.05}_{-0.07}) \times 10^{-13}$&$(9.2^{+2.7}_{-2.6}) \times 10^{-15}$\\
\hline
\end{tabular}
\begin{minipage}{16cm}
$^{a}$ The temperature was completely unconstrained for this
  model.
\end{minipage}
\end{table*}

It is likely that there is complicated temperature structure in the
shell that we cannot resolve, and as the gas in this region is assumed
to have passed through the shock front recently, it is likely that the
ionization balance has been affected by the shock. We fitted
non-equilibrium ionization models to the spectra for Region 3 (the
{\it nei} model in {\sc xspec}), and found a fit of similar quality to
that for the APEC model with a Gaussian at 1.35 keV, with a very
similar temperature and normalisation to our adopted model. Our
conclusions should not be strongly affected if the gas is far from
ionization equilibrium. It is likely that our spectral extraction
region contains gas with a range of temperatures and ionization
states; however, any effect this has on our conclusions below about
the gas density and shock dynamics is likely to be smaller than the
uncertainties introduced by our geometric assumptions in
Section~\ref{dyn}. In addition, the electron and proton fluids in the
shocked plasma may not be in equilibrium \citep[e.g.][]{kra07a}. We
discuss this possibility further in Section~\ref{secnoneq}.

As we are unable to constrain these effects with our data, we adopt
the simple, fixed abundance, single-temperature model with Gaussian
component to account for the prominent Mg line, as discussed above. We
adopt an abundance of $Z=0.5$ times solar to compute gas density and
pressure, rather than the lower value from the free abundance fit, as
this is likely to be artificially low due to using a single
temperature model, as discussed above. As the X-ray emissivity is due
to both continuum mainly from hydrogen and helium and line emission
from heavy elements and varies with the square of density, the
inferred gas density varies at most as $Z^{1/2}$. Thus, the assumed
abundance will introduce uncertainties of the order of ten per cent at
most.

\subsection{Spectral structure of the shell}
\label{spix}
\begin{figure*}
\centerline{\vbox{\hbox{
\epsfig{figure=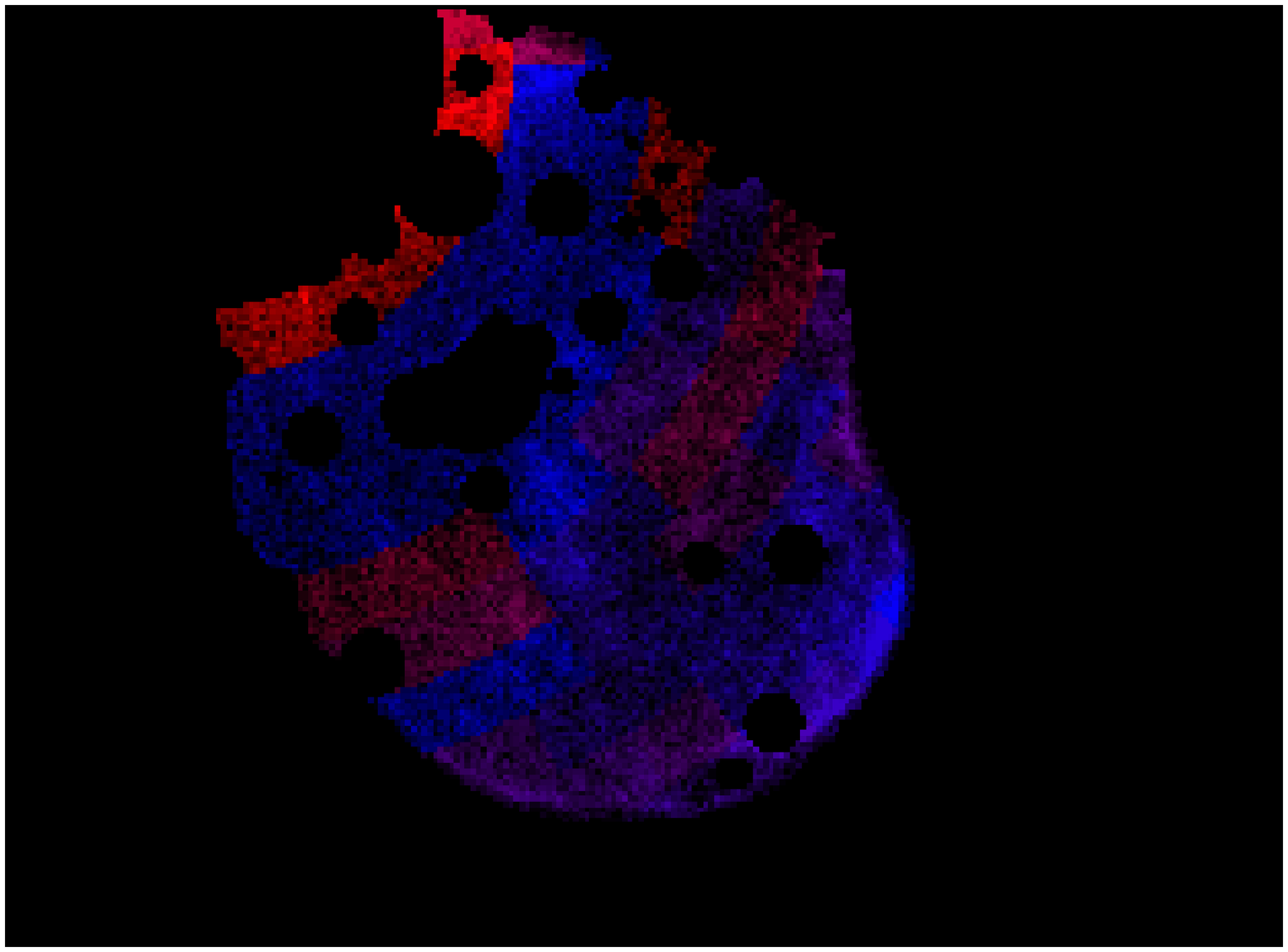,height=5.5cm}
\epsfig{figure=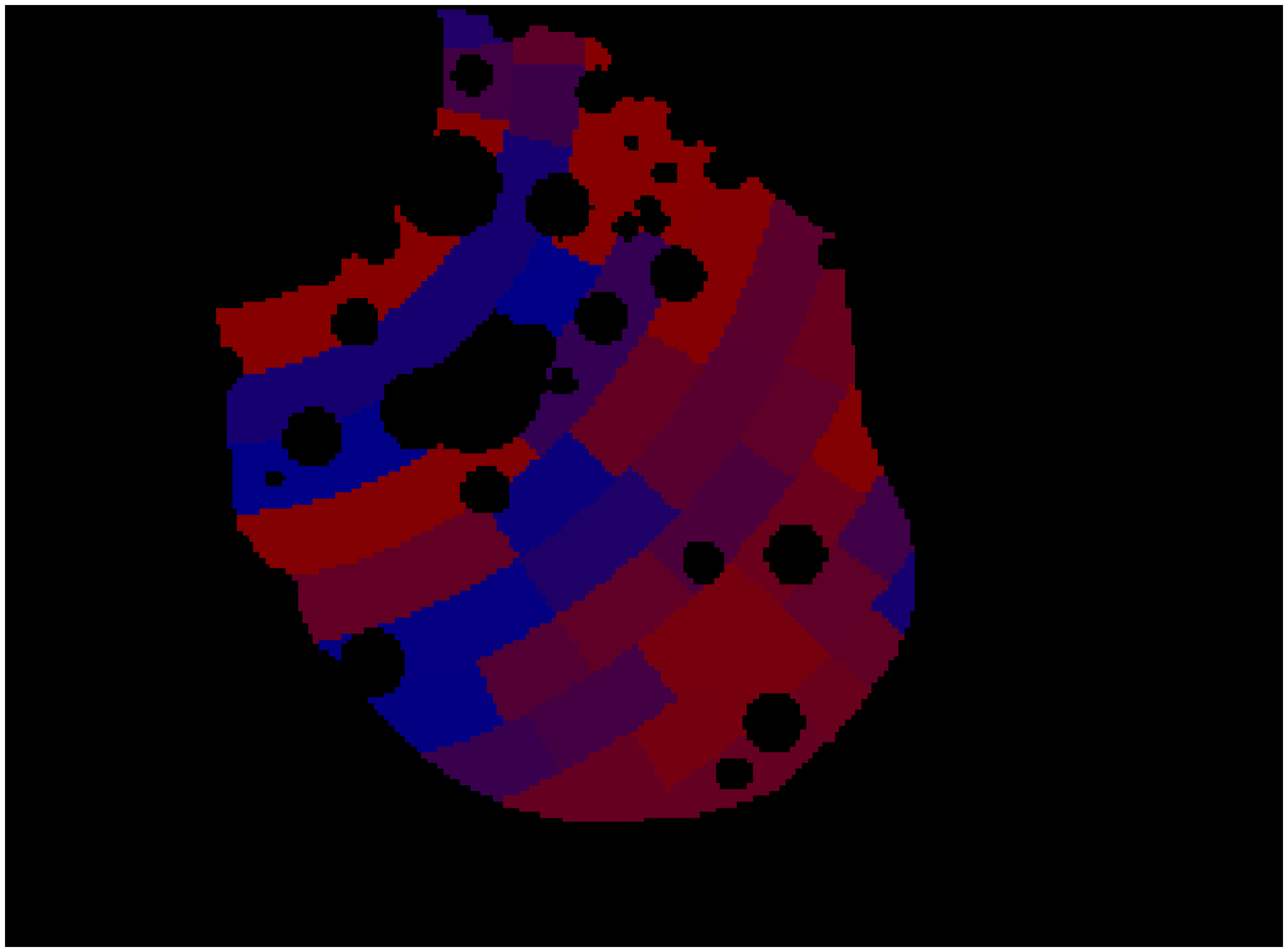,height=5.5cm}}
\hbox{
\epsfig{figure=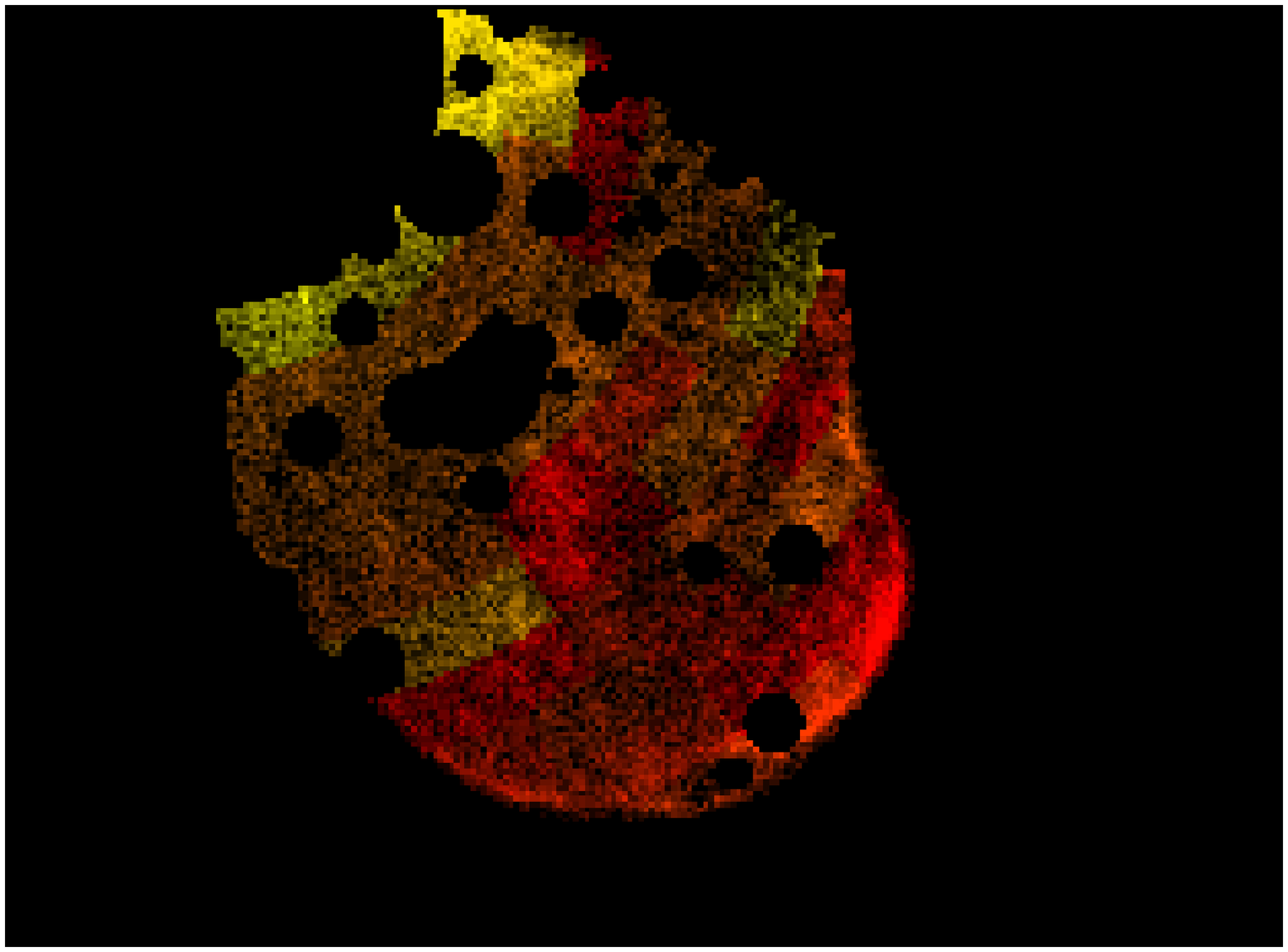,height=5.5cm}
\epsfig{figure=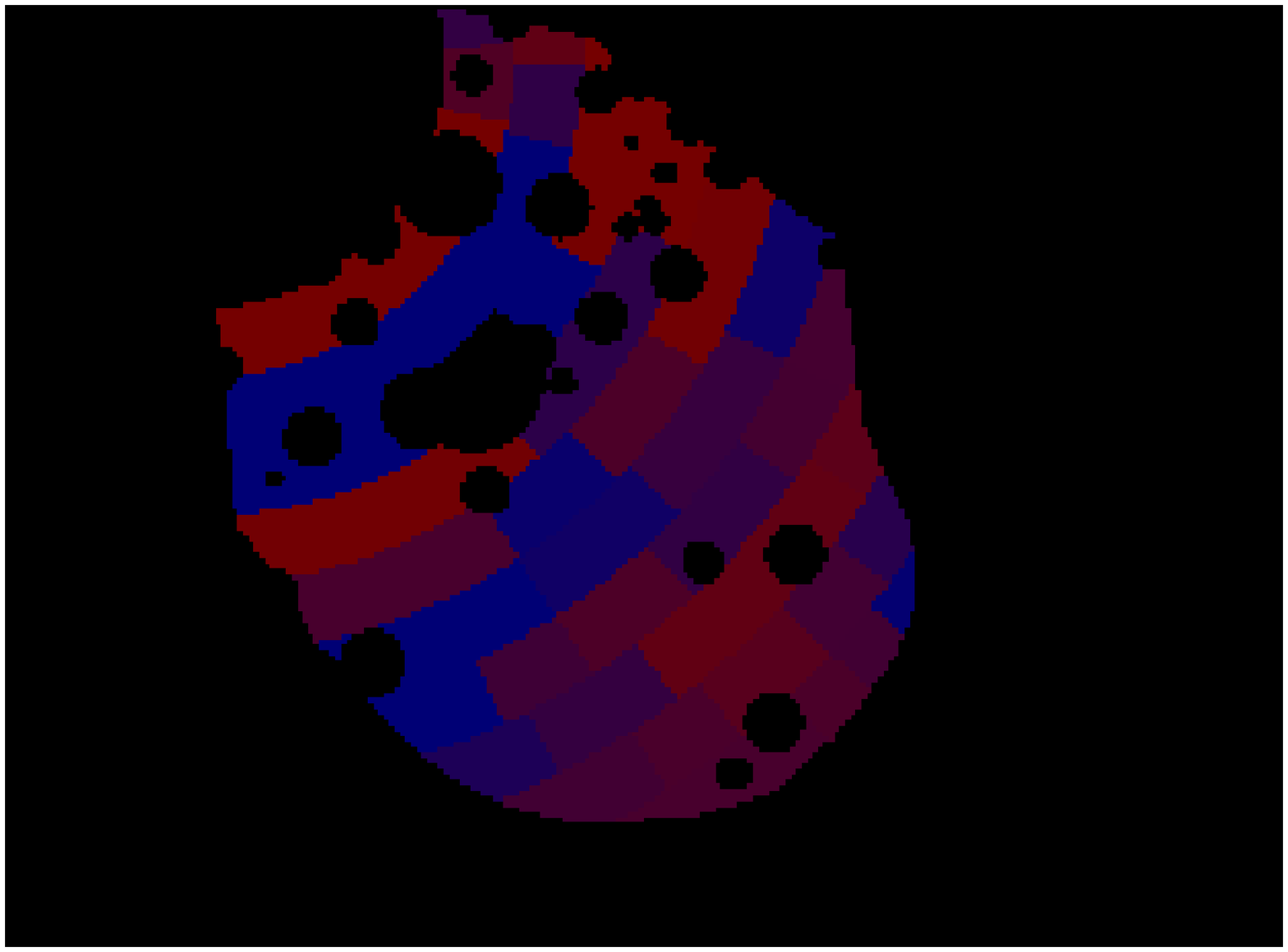,height=5.5cm}}
\vskip 0.7cm
\hbox{
\epsfig{figure=alpha_dist.ps,height=6cm}
\hskip 0.5cm
\epsfig{figure=fluxratio.ps,height=6cm}}}}
\caption{Spectral structure across the shell, as discussed in
Section~\ref{spix}. Top: spectral index map, assuming Model IV (power
law) for all regions (left: blue=1.5, red=4.0), with corresponding map
of $\chi^{2}/n$ (right: red = 0.5, blue=2.5). Middle: temperature map,
assuming Model III (APEC with $Z = 0.5$) for all regions (l: red=4.0
keV, yellow=0.7 keV), with corresponding map of $\chi^{2}/n$ (r: red =
0.5, blue=2.5). Bottom left: spectral index as a function of distance
from the nucleus. The large error bar for the third data point is
because a large fraction of this region is excluded due to a bright
X-ray transient present in several datasets (Sivakoff et al. 2008).
Bottom right: fraction of flux in thermal (red) and non-thermal (blue)
components for Model V as a function of distance from the nucleus.}
\label{spixmap}
\end{figure*}

We divided the entire lobe into small regions to map the spectral
structure across the shell. The regions were defined by requiring a
minimum of 2000 0.4 -- 2.5 keV counts (before background subtraction)
to allow good constraints to be obtained on the spectral parameters.
Background spectra were extracted from a region adjacent to the shell
at the same radius as the source region, in order to subtract
correctly the emission from the galaxy ISM and from the wings of the
nuclear PSF. This resulted in a total of 50 regions, which were each
fitted separately with APEC (Model III) and power-law (Model IV)
models in the same way as for the fits for larger regions described
above. Fig.~\ref{spixmap} shows a spectral index map for the power-law
fits (top panel), and a temperature map obtained using the APEC fits
(middle panel). For each spectral map, a corresponding map of the fit
statistic ($\chi^{2}/n$) is also shown.

As expected from our analysis above, the inner regions of the shell
(close to the AGN) have very different spectral properties to those in
the outer parts (e.g. Region 1) -- for the power-law fits, the inner
parts have much steeper spectral indices (and often very poor fit
statistics), consistent with a thermal interpretation for these
regions. In the region between $\sim 2$ arcmin and $\sim 3$ arcmin
from the nucleus none of the fits is good. This is partly because the
emission there has lower surface brightness than elsewhere, but mainly
due to contamination from the wings of the PSF from the bright
transient source at this radius \citep{siv08}. The outer parts are
best fitted by a power law, with the spectral index varying from $\sim
1.7$ to $\sim 2.4$, consistent with a predominantly synchrotron
origin.

The bottom left-hand panel of Fig.~\ref{spixmap} shows a plot of
spectral index as a function of distance from the central AGN. For
this analysis we defined a set of annuli centered on the nucleus, and
extracted spectra from regions consisting of the intersection of each
annulus with a polygon enclosing the shell region. We find that the
spectral index is steepest in the central regions, where the power-law
fit is not a good representation of the data. In the outer regions the
spectral index remains fairly close to a value of 2, but there is a
statistically significant decline in the spectral index in the last
three bins. This trend is present whether the results of Model IV or V
are adopted. This is discussed further in Section~\ref{dyn}.

We also investigated the contributions of thermal and non-thermal
emission to the observed flux as a function of distance from the
nucleus by fitting Model V (APEC + power law) to the same regions as
for the spectral index plot discussed in the previous paragraph. The
bottom right hand panel of Fig.~\ref{spixmap} shows the fraction of
the total unabsorbed 0.4 -- 3.0 keV flux in the thermal and
non-thermal model components as a function of distance from the
nucleus, indicating that the X-ray flux is dominated by non-thermal
emission except in the very inner regions. The strong evidence for
dominance by thermal emission in Region 2 leads us to conclude that
this is likely to be a real difference in the shell emission, rather
than an artefact of poor background subtraction.

\section{Discussion}
\label{discuss}
Our spectral analysis of the X-ray shell emission strongly favours a
non-thermal interpretation for the majority of the X-ray emission in
the outer half of the shell, due to the unrealistically low abundances
required by thermal models, as well as the lack of expected
emission-line signatures, particularly the Fe L shell complex. These
results, together with the strong resemblance between the highly
edge-brightened shell emission and the X-ray emission from
synchrotron-dominated supernova remnants, such as SN1006
\citep[e.g.][]{rot04}, lead us to conclude that an X-ray synchrotron
origin for the majority of the X-ray shell emission from Cen A is
viable. Further constraints on this model come from limits at other
wavelengths. It might be expected that a plausible energy distribution
for an electron population producing X-ray synchrotron emission would
also produce detectable emission at radio, optical, UV or infrared
wavelengths, and so the model must pass this test. It must also be
possible to construct a self-consistent picture of the radio-lobe
dynamics, which explains both the strong thermal emission from Region
3 and the lack of any thermal signatures in the outer rim region.
Below we discuss the constraints on a synchrotron model that can be
obtained from multiwavelength measurements and from an analysis of the
shock conditions and lobe dynamics. We also consider the dependence of
our conclusions on the assumption of electron-ion equilibrium.
Finally, we consider the implications of our results for particle
acceleration in Cen A and other radio galaxies, and for Cen A as a
source of ultra-high energy cosmic rays (UHECRs) and TeV emission.

\subsection{The broad-band SED of the shell region}
\label{sed}
\begin{figure*}
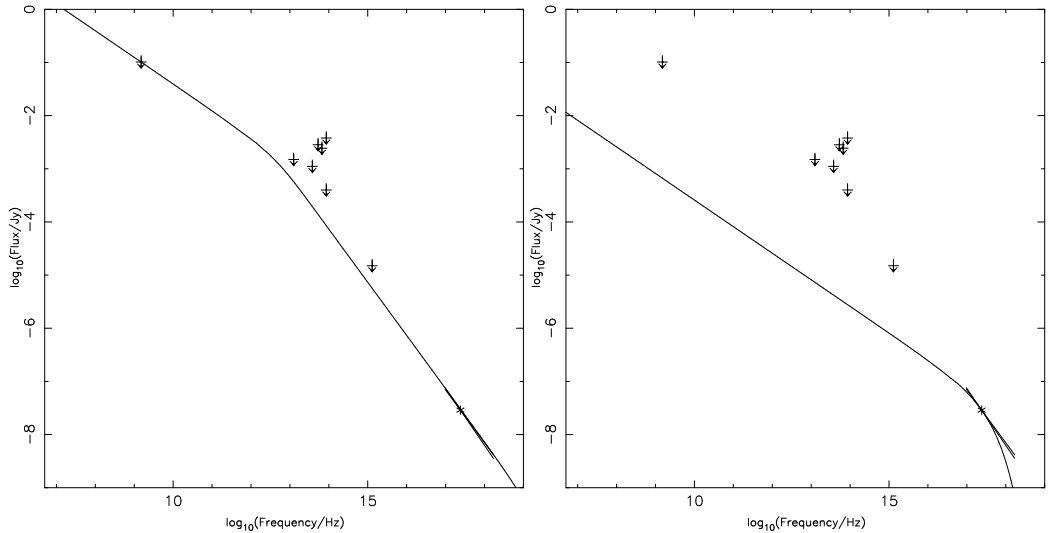

\begin{center}
\centerline{\hbox{
\epsfig{figure=cena_shell_sed.ps,height=7cm}
\epsfig{figure=cena_shell_sed2.ps,height=7cm}}}
\caption{The broadband SED for Region 2, including upper limits at
  1.4-GHz, 3.6, 4.5, 5.8, 8.0 and 24 $\mu$m, 231 nm, and the measured
  X-ray flux density at 1 keV and spectral index. Solid line is a
  synchrotron model fit with (l) $\alpha = 0.5$, $\gamma_{min} = 10$,
  $\gamma_{break} = 4.4 \times 10^{4}$, $\gamma_{max} = 5 \times
  10^{9}$ (discussed in Section~\ref{sed}), and (right panel) $\alpha
  = 0.5$, no break, and $\gamma_{max} = 3 \times 10^{8}$ (discussed in
  Section~\ref{implic}).}
\label{sedfig}
\end{center}
\end{figure*}

The brightest part of the X-ray shell surrounding Cen A's SW lobe is
the outer rim, which is spatially separated from the region coincident
with the radio emission from the radio lobe itself. If the X-ray
emission from the shell is produced via the synchrotron process, then
it would be expected to originate in an electron population that
extends to energies much lower than those emitting in the X-ray band.
No shell emission has been reported previously at radio, optical, IR,
UV or radio wavelengths. We determined limits in the radio, infrared
and UV using VLA, {\it Spitzer} and {\it GALEX} observations to
construct a broadband SED (Fig.~\ref{sedfig}) for a region of shell
emission surrounding the radio lobe (we used Region 2, discussed
above, which is exterior to the radio lobe in projection, so as to
obtain as tight a radio constraint as possible).

The radio limit shown in Fig.~\ref{sedfig} was obtained from the
1.5-GHz radio map described in Section~\ref{analysis} and is a $3\sigma$
limit based on the off-source rms noise, as there is no detection of
radio emission from Region 2. We used the {\it Spitzer} IRAC and MIPS
\citep{bro06,mjh06} and GALEX observations \citep{nef03,mjh06} to
obtain infrared limits at 3.6, 4.5, 5.8, 8.0 and 24 $\mu$m and a UV
limit at 231 nm. The analysis was carried out as described in
\citet{mjh06}. Limits were calculated for Region 2, using a larger
region immediately outside Region 2 as a background. At all bands from
the IR to the UV the dominant background is stars, and so this choice
of background will give a strict upper limit. As an alternative
strategy, we also used galaxy-subtracted IR images to obtain tighter
limits on the IR emission from the shell. This allowed us to decrease
the upper limit at 3.6 $\mu$m by roughly an order of magnitude.
Similar improvements could be obtained using this method at other
wavelengths, but would not significantly affect the constraints on a
synchrotron model, as discussed below. Figure~\ref{sedfig} includes all
of the limits obtained, including the galaxy-subtracted limit at 3.6
$\mu$m.

Fig.~\ref{sedfig} shows a synchrotron model chosen to fit through both
the X-ray data point and spectral index and the 1.4-GHz upper limit.
The plotted model assumes $\alpha = 0.5$, $\gamma_{min} = 10$,
$\gamma_{break} = 4.4 \times 10^{4}$, $\gamma_{max} = 5 \times
10^{9}$, and uses the 1.4-GHz upper limit to normalise the electron
spectrum, assuming equipartition in the shell material. The volume
enclosed in Region 2 was modelled by approximating to a box of
dimensions $573 \times 7 \times 46$ arcsec, where the length and width
are the dimensions of the extraction region, and the depth is
approximated by that halfway between the inner and outer curved
boundaries of the regions, assuming a spherical shell of appropriate
radius to fit the shell curvature. This gives a volume of $\sim 3
\times 10^{64}$ cm$^{3}$. None of the infrared or UV limits can rule
out the synchrotron model described above. In addition, since the
radio data point is an upper limit, a model with a significantly
higher break frequency (as typically found for supernova remnants with
X-ray synchrotron emission, e.g. \citealt{dye04,rho03}; see also
Section~\ref{implic}), which could be orders of magnitude below the IR
limits, also cannot be ruled out (see the right hand panel of
Fig.~\ref{sedfig}). Hence we conclude that the broadband SED for the
rim of the Cen A SW shell is consistent with a synchrotron
interpretation for the X-ray emission.

Although a non-thermal contribution to the spectrum in the NE part of
the shell (Region 3) is not required, the limits on a power-law
contribution to the Region 3 spectrum from fitting Model V are
consistent with a significant flux from synchrotron emission in this
region (the fit for Model V implies an unabsorbed 1-keV flux density
from the power-law component of $\sim 2$ nJy). Therefore, we cannot
rule out significant particle acceleration in Region 3 as well.

Finally we also investigated whether inverse Compton emission could
explain some or all of the shell X-ray emission. X-ray emission from
IC scatting could be significant in the regions coincident with the
radio lobes (whose X-ray emission we have hitherto been interpreting
as coming from the shell material in front of and behind the lobe). We
therefore modelled the electron distribution in the radio lobe in
order to predict the expected level of IC emission. We approximated
the lobe as a cylinder of length 270 arcsec and radius 80 arcsec,
using radio flux measurements at 240 MHz and 8 GHz to determine the
shape and normalisation of the electron distribution, assuming
equipartition. We assumed $\gamma_{min} = 10$, as above, and chose
$\gamma_{max}$ to fit the observed radio spectrum, giving a value of
$1.2 \times 10^{4}$. The predicted 1-keV flux density for the lobe
from inverse-Compton scattering of the cosmic microwave background
radiation (IC/CMB) (the dominant process) and synchrotron self-Compton
(SSC) is $\sim 3.3$ nJy, which is factor $\sim 30$ below the observed
flux from the lobe region. Hence it is possible that a small fraction
of the observed X-ray flux from Region 1 is produced by the IC/CMB
process. For the lobe emission to be produced entirely by IC/CMB would
require a magnetic field a factor $\sim 6$ below equipartition. An
inverse-Compton model for the edge-brightened rim of the X-ray
structure (e.g. Region 2) is implausible, due to our tight limits on
the radio emission in this region, and because none of the many
examples of IC/CMB-detected lobes \citep[e.g.][]{jhc05} show
edge-brightening. We therefore consider it unlikely that our
conclusions are affected by possible contamination from X-ray IC
emission.

\subsection{Lobe expansion and shock dynamics in a synchrotron model}
\label{dyn}
As the X-ray emission from the north-east region of the shell (Region
3) is unambiguously thermal in origin (see Section~\ref{s1}), this is
the best region to use for investigating the shock jump conditions. We
approximated the volume of shell enclosed within Region 3 as a box of
dimensions $0.480 \times 0.127 \times 0.769$ arcmin, where the length
and width are the dimensions of the spectral extraction region, and
the depth along the line of sight was taken to be the chord of a
circle having the curvature of the outer edge of the shell in the
north-east region at a distance halfway between the inner and outer
edges of the spectral extraction region. This leads to an assumed
volume of $1.7 \times 10^{63}$ cm$^{3}$. With this volume, and
assuming the APEC + Gaussian model described in Section~\ref{s1}, we
obtain a density of $n_{p} = 0.033$ cm$^{-3}$, which corresponds to a
pressure in the shocked gas of $1.1 \times 10^{-10}$ dyne cm$^{-2}$,
assuming $kT = 0.95$ keV as discussed above. The shell density in
Region 3 is roughly 3.5 times the density of the ISM at this distance,
based on the beta model fitted to the external gas density
distribution by \citet{kra03}, although we note that the ISM density
is not well constrained, because the distribution of gas is likely to
deviate significantly from spherical symmetry, and to be inhomogeneous
in the inner regions of the galaxy. Hence the density constrast is
consistent with the value of 4 expected for a strong shock, as was
concluded previously by \citet{kra07a}. If the hot-gas shell and radio
lobe are assumed to be in pressure balance, then the pressure inside
the radio lobe is expected to be $\sim 1.1 \times 10^{-10}$ dyne
cm$^{-2}$, as calculated above, which is about an order of magnitude
higher than the lobe minimum pressure. This is of the same order as
the differences between lobe minimum pressures and external pressures
for FRI radio galaxies in general \citep[e.g.][]{jhc07,jhc08}. The
pressure contrast between the shell and the ISM is $P_{shell}/P_{ism}
= 10$, using the gas density distribution and temperature from
\citet{kra03}. Assuming ram-pressure balance and that $\gamma=5/3$,
this pressure jump implies that, at its north-east edge, the radio
lobe is expanding at a Mach number of $\sim 2.8$, which corresponds to
an expansion speed of $\sim 850$ km s$^{-1}$.

\begin{table}
\caption{Adopted physical parameters for Regions 1 and 3}
\label{params}
\begin{tabular}{lrr}
\hline
Parameter&Region 2&Region 3\\
\hline
$T_{shell}$ (keV)&11.0&0.95\\
$T_{ism}$ (keV)&0.35&0.35\\
$Z$ (Z$_{\sun}$)&0.5&0.5\\
$n_{p,shell}$ (cm$^{-3}$)&0.004&0.033\\
$n_{p,ism}$ (cm$^{-3}$)&0.001&0.01\\
$P_{shell}$ (dyne cm$^{-2}$)&$1.1 \times 10^{-10}$&$1.1 \times 10^{-10}$\\
$P_{ism}$ (dyne cm$^{-2}$)&$1.3 \times 10^{-12}$&$1.1 \times 10^{-11}$\\
\hline
\end{tabular}
\end{table}
\citet{kra03} found that the apparent density contrast between shocked
and unshocked gas in the outer parts of the shell (assuming a thermal
interpretation for all the shell X-ray emission) is much higher than
the prediction for a strong shock, and derived the expansion speed
assuming that the observed shell is shocked gas that has cooled rather
than the component that directly enters the jump equations for shocked
gas. A non-thermal model for most of the X-ray emission within the
shell can provide a simpler interpretation of the larger
X-ray-intensity constrast inside and outside the shell. We checked the
low apparent contribution from thermal emission in the outer regions
of the shell is consistent with the thermal emission expected from a
shell of shocked gas a factor of 4 denser than the ISM in the shell's
outer regions (for several assumed temperatures). We assumed an
unshocked ISM density of $1.0 \times 10^{-3}$ cm$^{-3}$ at the outer
edge of the shell (Kraft et al. in prep.: note that this is a factor
1.7 lower than that assumed by \citealt{kra03} and \citealt{kra07b}),
so that the shell gas would be expected to have a density of $\sim 4
\times 10^{-3}$ cm$^{-3}$. We calculated the volume corresponding to
the outer shell region discussed in Section~\ref{outer} assuming a
half spherical shell of outer radius 121 arcsec. We estimated the
thickness of the shell using the 1.4-GHz radio map and the merged
0.4--2.5 keV event file by measuring the separation between the edge
of the radio lobe and the brightest part of the shell edge. We took
the average of three measurements around the edge of the shell, giving
a mean thickness of $16.3\pm1.5$ arcsec, where the measurement
uncertainty (which may be partly due to real variations in the
thickness of the shock front) is larger than any astrometric
uncertainty or resolution effects. Our measured shell thickness is
consistent with the value of $17$ arcsec adopted by \citet{kra07a}.
This assumed geometry leads to an expected emission measure of $\sim
1.7 \times 10^{58}$ cm$^{-3}$ (corresponding to a 0.4 - 3.0 keV flux
of $1.3 \times 10^{-14}$ erg cm$^{-2}$ s$^{-1}$). The measured
normalization for the APEC component of Model V discussed in
Section~\ref{outer} corresponds to an emission measure more than
thirty times higher, hence the data are consistent with this emission
being present. However, as the APEC component in the spectrum of this
region is not required by the data, the spectra can also be well
fitted by an APEC + power-law model with the APEC normalization fixed
at the value corresponding to the expected emission measure, for any
reasonable temperature (e.g. between $kT = 0.5$ and $15.0$ keV -- our
adopted shock model as discussed below implies a post-shock
temperature of $\sim 11$ keV). In all cases the thermal contribution
to the fit is negligible. Hence the X-ray spectra for the outer shell
are consistent with a model in which a strong shock is occurring, with
the X-ray emission dominated by synchrotron emission from high-energy
particle acceleration.

If we assume that the radio lobe is isobaric, and that the radio lobe
and shocked gas are in pressure balance everywhere, then the pressure
in the shocked gas shell at its outer edge (e.g. Region 2) should be
the same as the lobe internal pressure and shell pressure in Region 3,
which we calculated above. This implies that the expansion speed at
the outer edge of the lobe where we see the brightest non-thermal
emission should be much higher, given the significantly lower external
gas density ($\sim 10^{-3}$ cm$^{-3}$). Assuming an ISM gas
temperature of 0.35 keV (Kraft et al. in prep), we obtain a pressure
jump $P_{shell}/P_{ism} \sim 87$, corresponding to a Mach number of
$\sim 8.4$, and a lobe expansion speed at its outer edge of $\sim
2600$ km s$^{-1}$. This speed is similar to that inferred by
\citet{kra03}. We can also infer a transverse expansion speed for the
radio lobe similarly, using the external density at the lobe midpoint
given by the profile in \citet{kra03}, which we find to be $\sim
1400$ km s$^{-1}$. The observed axial ratio of the radio lobe is
roughly consistent with the inferred longitudinal and transverse
expansion speeds for the lobe if the expansion has been self similar
throughout the source's lifetime (though in practice this is unlikely
to have been the case). The adopted physical parameters for Regions 2
and 3 are summarized in Table~\ref{params}.

As for Region 3, we have assumed here that $\gamma=5/3$. This will be
correct if the shell pressure is dominated by the thermal component of
the shocked gas, rather than by the pressure in the relativistic
particles that are producing the observed X-ray synchrotron emission.
As we concluded above, the observations are consistent with a scenario
in which the hot shocked gas has the density expected for a strong
shock, but is not separable from the dominant emission attributed to
synchrotron radiation. However, we also checked whether the internal
pressure of the synchrotron-emitting particles is likely to
significantly alter the shock dynamics. Assuming the SED discussed in
Section~\ref{sed} with an equipartition magnetic field of $\sim 8$ $\mu$G
and no energetically significant protons, we find that the pressure in
cosmic rays and magnetic field is $\sim 2 \times 10^{-12}$ dyne
cm$^{-2}$, which is insignificant compared to the thermal pressure. We
note that this is an upper limit on the pressure in the relativistic
particles if in equipartition, because the SED model is based on upper
limits at radio, IR and UV wavelengths. Magnetic field strengths of
$\sim 90 \mu$G or higher, more than an order of magnitude higher than
the equipartition value, would be required for the non-thermal
(magnetic) pressure to dominate, assuming $\gamma_{min} = 10$. A
departure from equipartition in the direction of electron dominance
could also lead to non-thermal pressure dominating over thermal
pressure; however, this would also have to be extreme.

We can also estimate the total energy of the radio source and the
power of the jet. Approximating the shock front as a prolate spheroid
with semi-major axis 167.2 arcsec and semi-minor axis 115.0 arcsec, we
find an enclosed volume of $1.6 \times 10^{66}$ cm$^{3}$. The cavity
enthalpy, $4PV$ is then $7 \times 10^{56}$ ergs. We make a rough
estimate of the kinetic energy of the shell using the expansion speed
midway along the shell, and the gas density for Region 1, to obtain
$E_{K} = 1.2 \times 10^{56}$ ergs. Assuming a constant expansion
speed, we obtain a source age estimate of $\sim 2 \times 10^{6}$ y,
and hence we estimate a mean jet power over the source lifetime of
$\sim 10^{43}$ erg s$^{-1}$. We can also estimate an instantaneous jet
power as $P\dot{V}$, where $\dot{V} = V(\dot{a}/a + 2 \dot{b}/b)$,
with $a$ and $b$ the semi-major and semi-minor axes respectively. If
the lobe expansion is self-similar, then $\dot{V} = 3V\dot{a}/a = 4
\pi b^{2} \dot{a}$. As one end of the expanding lobe is tied to the
AGN, $2\dot{a} = v_{shock}$, the shock expansion speed at its outer
edge, and hence the instantaneous jet power, $p$, is given by $p = 2
\pi P b^{2} v_{shock} = 6.6 \times 10^{42}$ erg s$^{-1}$ (using the
shock speed of 2,600 km s$^{-1}$ at the outer edge of the shock). This
is roughly a factor 15 times higher than the current radiative power
of the AGN \citep{eva04}.

\subsection{Electron-ion equilibrium in the shocked gas}
\label{secnoneq}

\begin{figure*}
\centerline{\hbox{
\epsfig{figure=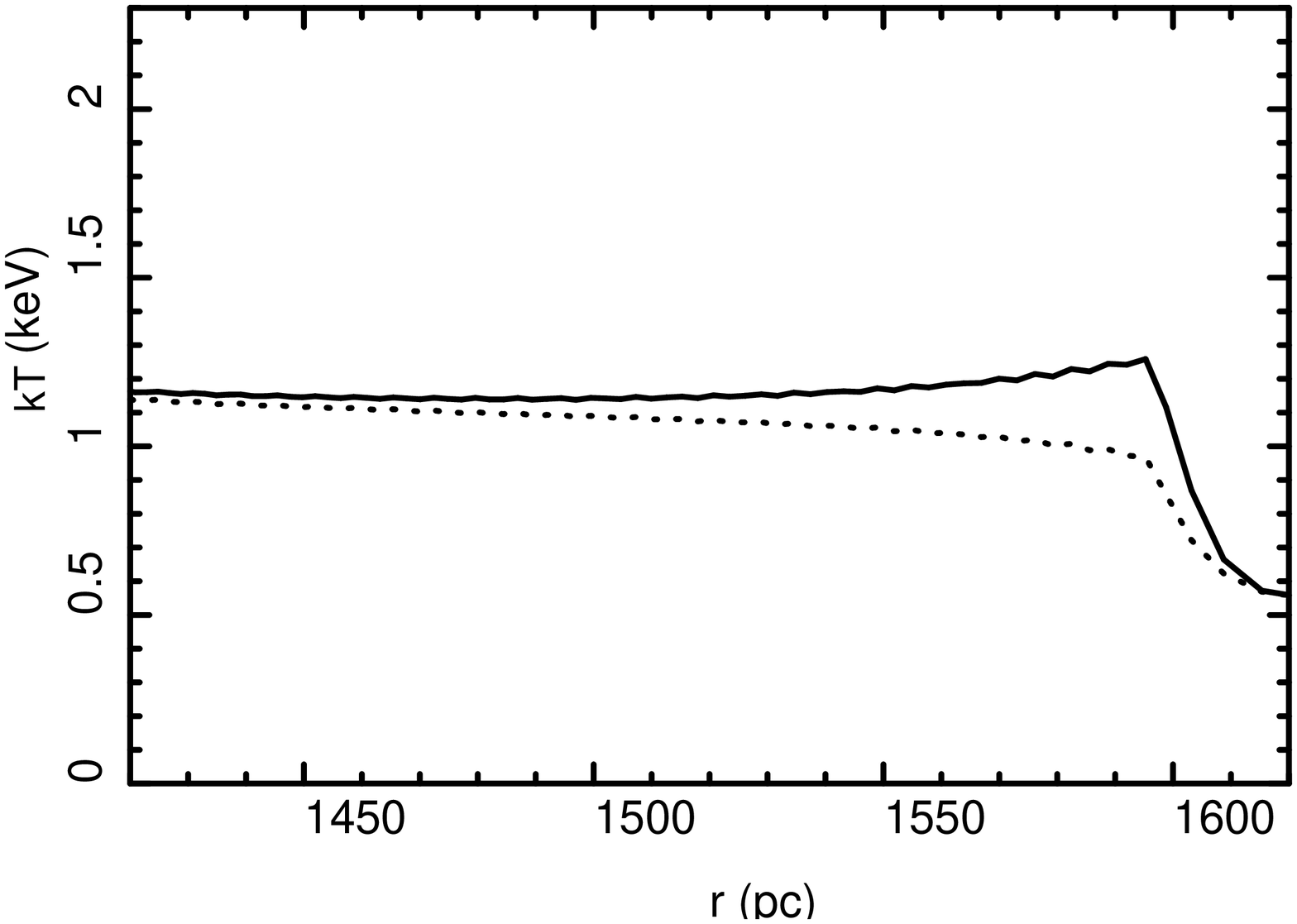,height=5.5cm}
\epsfig{figure=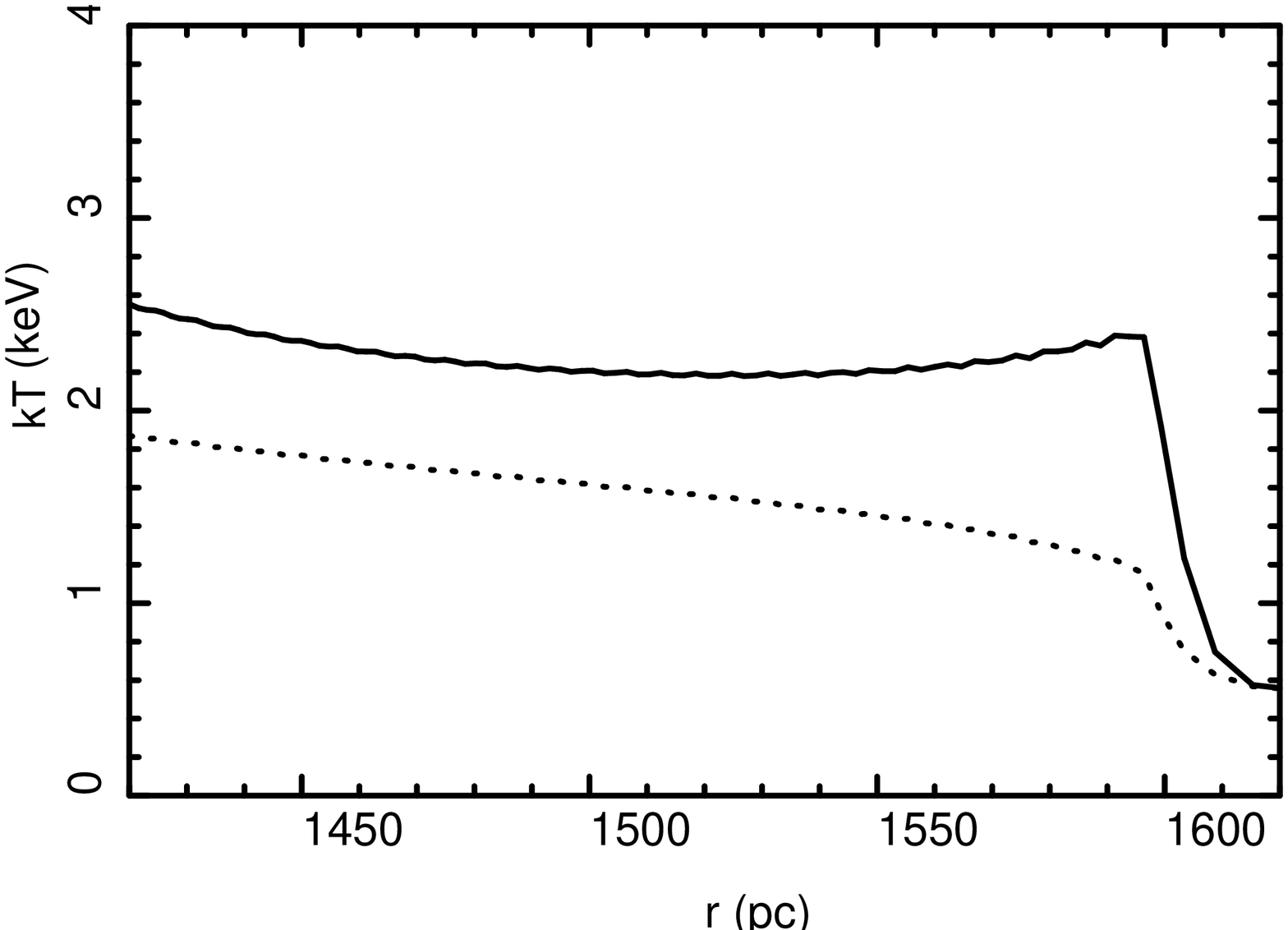,height=5.5cm}}}
\caption{Post-shock proton temperature (solid line) and electron
  temperature (dotted line) profiles obtained from our two-fluid
  hydrodynamical modelling (Section~\ref{secnoneq}) for $kT = 1$ keV (left)
  and $kT = 1.4$ keV (right).}
\label{noneq}
\end{figure*}

Our conclusions about the shock dynamics in Section~\ref{dyn} depend
strongly on our adopted spectral model for the thermal emission in
Region 3. As discussed in Section~\ref{s1}, the region shows a somewhat
unusual spectrum, with a prominent magnesium line, that is not well
fitted by a single APEC model. As the gas in this region is assumed to
have passed through the shock front recently, it is likely that the
ionization balance has been affected by the shock, as discussed in
Section~\ref{s1}. It is also likely that our spectral extraction region
contains gas with a range of temperatures and ionization states, but
in Section~\ref{s1} we concluded that any effect on our conclusions about
the gas density and shock dynamics is likely to be smaller than the
uncertainties introduced by our geometric assumptions in Section~\ref{dyn}.
However, an additional complication is that the electron and ion
fluids may not have achieved equilibrium, as was discussed in the
context of earlier work on the Cen A shock by \citet{kra07a}.

For gas temperatures of $\sim 1$ keV and below, and photon energies
below $\sim 1$ keV, at the energy resolution of ACIS-I the thermal
continuum is submerged by lines, particularly those of the Fe L
complex. This makes fitted temperatures insensitive to the continuum.
If the gas is not in ionization equilibrium, the relationship between
line strengths and gas temperature is model-dependent, so that a
detailed model of the shock and subsequent return to ionization
equilibrium is required to determine the gas temperature accurately.
On the other hand, the shape of the continuum is determined largely by
the electron temperature. In the absence of a detailed model for the
ionization state of the gas, we have attempted to constrain the
postshock electron temperature by fitting the spectrum of Region 3 in
the energy range 1.5 -- 3 keV, where the lines are weaker relative to
the continuum. A 1.3 keV Gaussian line was included in this fit, as
discussed in Section~\ref{s1}. Its energy and normalization were fixed by
fitting it together with an APEC model over the energy range 0.5 -- 2
keV. For the 1.5 -- 3 keV fit, the abundances were fixed at 0.5 solar.
The resulting 90\% confidence range for the temperature is 0.52 --
1.39 keV, consistent with the previous thermal fit.

If the flow time through Region 3 is comparable to or smaller than the
electron-ion equilibration time, then the shocked protons there could
be significantly hotter than the electrons \citep[cf.][]{kra07a}. In
that case, the postshock pressure may be significantly greater than
estimated above, only providing a lower limit on the true pressure of
the radio lobe. The value of the lobe pressure is important, since it
determines the strength and speed of the shock, and the estimated jet
power scales approximately as $P^{3/2}$, so this issue warrants
careful scrutiny.

To determine how much the electron and proton temperatures may differ,
we have used the two-fluid hydrodynamic model of \citet{kra07a}.
Motivated by observations of Galactic supernova remnants, this model
assumes that the protons are heated rapidly in a collisionless shock,
while the electrons are heated by Coulomb collisions with the protons.
The model is spherically symmetric and, for the cases illustrated
here, the shock is driven by constant power input into an expanding
``radio lobe'' at the centre of an initially uniform atmosphere. For
scaling purposes, the shock radius is taken to be 1.6 kpc (90 arcsec).

In terms of the electron and proton temperatures, $T_{\rm e}$ and
$T_{\rm p}$, respectively, the energy exchange can be expressed as
$dT_{\rm e}/dt = \nu (T_{\rm p} - T_{\rm e})$, where the rate
coefficient is determined chiefly by the electron temperature,
approximately $\nu \propto n_{\rm e} T_{\rm e}^{-3/2}$
\citep[e.g.][]{kra07a}. Although the electron-ion equilibration time
is relatively short, $\simeq 2\times10^5$ y for $n_{p} = 0.033\rm\
cm^{-3}$ and $kT = 0.97$ keV, it is not much shorter than the sound
crossing time of Region 3, $\simeq 2.6\times10^5$ y. Increasing the
temperature of the shocked gas increases the electron-proton
equilibration time while also decreasing the sound crossing time.
Thus, although electrons and protons in Region 3 should have much the
same temperature if the electron temperature there is close to 1 keV,
their temperatures can differ significantly if the electron
temperature is moderately higher.

This is illustrated in Fig.~\ref{noneq}, which shows postshock proton
and electron temperature profiles for two cases. In the first case
(Fig.~\ref{noneq}a), the emission measure weighted electron
temperature in Region 3 (1.48 -- 1.60 kpc) is 1 keV and the RMS proton
density is $0.033\rm\ cm^{-3}$. In this case the difference between
the electron and proton temperatures is negligibile. For the second
case (Fig.~\ref{noneq}b), the RMS density is the same, but the
emission measured weighted electron temperature is close to 1.4 keV,
the 90 per cent level upper limit. Here, the proton temperature is
$\sim70$ per cent greater than the electron temperature. The postshock
pressure would then be roughly 35 per cent larger than suggested by
the electron temperature. This disparity between electron and proton
temperatures grows rapidly with electron temperature. However, if the
electron temperature is smaller than the 1.4 keV upper limit, our
estimate for the postshock pressure should be a reasonable estimate of
the true pressure.

\subsection{Implications of the X-ray synchrotron interpretation for
  particle acceleration at radio-lobe shocks}
\label{implic}
In Section~\ref{sed} we demonstrated that a synchrotron origin for the
shell X-ray emission is consistent with the broad-band SED of the
shell emission in a region exterior to the radio lobe (where the most
constraining radio limits could be obtained). We have also shown that
the lack of thermal X-ray emission from the outer parts of the shell
is consistent with the expected level of emission from a strong shock
-- the observed thermal X-ray emission in the more central NW shell
region (Section~\ref{s1}) and the lack of any signatures of such emission
in its outer parts (Section~\ref{outer}) are as expected for a strong
shock propagating into Cen A's environment. We conclude that the
majority of the shell X-ray emission is synchrotron emission from
high-energy particle acceleration at the shock front. Interestingly,
the inferred expansion speed of the radio lobe at its outer edge, as
calculated in the previous section, is very similar to the expansion
speeds of supernova remnants (SNRs) that show X-ray synchrotron
emission \citep[e.g.][]{vin06,war05,rot04,ace07}, whereas at the NW
edge, where no significant non-thermal emission is detected (although
particle acceleration to X-ray-emitting energies cannot be ruled out
-- see Section~\ref{sed}), the expansion speed is significantly lower. If
the inner lobes of Cen A are indeed expanding at these high
velocities, then it is perhaps not unexpected to see such a strong
signature of high-energy particle acceleration.

The statistically significant decrease in spectral index in the outer
$\sim 50$ arcsec of the shell (Section~\ref{spix}) could indicate
differences in the particle acceleration behaviour as a function of
distance and hence of shock expansion velocity. A possible
interpretation is that the spectrum is flattest at the outermost edge
of the shell (where the expansion speed is highest) because of a
higher value for $\gamma_{max}$, the observed high-energy cut-off of
the electron population, with the high-energy cut-off moving to lower
energies, thus steepening the spectrum, in regions closer to the
nucleus where the shock front is slower due to a denser environment.
This would require $\gamma_{max}$ in the range $10^{7} - 10^{9}$ over
the bright shell region, and significantly lower than $7 \times
10^{7}$ in Region 3. (We note that, assuming the observed value of
$\gamma_{max}$ is determined by radiative losses, the high-energy
cut-off for hadrons would be significantly higher). If the
X-ray-emitting electrons are being accelerated by diffusive shock
acceleration, and if the high-energy cut-off is determined by the
synchrotron loss timescale, then we can determine the dependence of
$\gamma_{max}$ on shock speed, as well as the expected value of
$\gamma_{max}$ for the shock conditions we observe. The high-energy
cut-off for particle acceleration has been discussed extensively in
the context of X-ray synchrotron emission from supernova remnants
\citep[e.g.][]{lc83,rey96}. For electrons, the acceleration timescale
to a given energy $E$ depends on $B_1$ (the pre-shock magnetic field
strength), $E$, $v_{shock}$, the ratio of the electron mean free path
along the magnetic field to the gyroradius ($f$, assumed to between 1
and $\sim 10$), and on a factor $R_{J}$ that takes into account the
decrease in efficiency of scattering if the magnetic field direction
is not parallel to the shock. We used equation (1) of \citet{rey96}:
\begin{equation}
E_{max} \approx 0.1 (f R_{J} B_{1})^{-1/2} v_{8} {\rm \ \ \ ergs}
\end{equation}
where units are cgs and the shock speed is in units of $10^{8}$ cm
s$^{-1}$. We assumed $f = 10$, $R_{J} = 1$ and $B_1 = 1 \mu$G, and
find that $\gamma_{max} \sim 10^{8}$ for the outer edge of the shell
and $\sim 3 \times 10^{7}$ at Region 3. In order to fit a synchrotron
model consistent with the observed X-ray spectral index for Region 2,
and with equipartition $B$ fields, we require $\gamma_{max}$ in the
range $2 \times 10^{8} - 4 \times 10^{8}$ (see Fig.~\ref{sedfig}),
which is roughly consistent with this analysis. This suggests that a
model in which the observed X-ray spectral changes can be explained by
changes in $\gamma_{max}$ as a function of shock speed is viable. We
have not considered the role of changes in $B_1$ and $f$ along the
shell in this discussion -- these are likely to vary somewhat with
position as well, but the dependence of $\gamma_{max}$ on these
parameters is weaker so that the change is shock speed is likely to be
the dominant effect.

\subsection{Implications for large-scale shocks in other galaxies}

If we accept a synchrotron origin for the X-ray emission from
Cen A, then the X-ray properties of the outer shell region
cannot be used to measure the impact of the radio galaxy on its
environment directly; however, the detection of thermal emission from
the shocked gas in Region 2 allows us to infer the dynamics of the
shocked shell and radio lobe indirectly as discussed above. The
results on the overall dynamics of the radio lobe and its
environmental impact are not significantly different from the
conclusions of \citet{kra03}. Unlike the shock in Cen A, the
strong shock detected in the similar system NGC\,3801 \citep{jhc07} is
clearly dominated by thermal X-ray emission, with no significant X-ray
synchrotron emission. It is likely that the lobes of NGC\,3801 are
expanding much more slowly than the outer parts of Cen A ($\sim
750$ km s$^{-1}$, similar to Cen A's expansion speed at Region 3), and
so the lack of strong high-energy particle acceleration in that system
is probably explained by this lower speed.

The bolometric X-ray luminosity of the shell emission in Cen A is
$\sim 4 \times 10^{39}$ erg s$^{-1}$, and so shocks on similar
physical scales will be undetectable in more distant radio galaxies;
however, these results could also have implications for powerful FRII
radio galaxies, whose lobes are much larger. A number of nearby FRII
radio galaxies have been found to reside in comparatively gas poor
environments \citep[e.g.][]{kra05,kra07b}, and so may be having
similarly dramatic effects on their environments. If supersonic
expansion is occurring in FRIIs at speeds similar to that of Cen A,
then high-energy particle acceleration could be important in FRII
environments. We considered the example of the nearby FRII radio
galaxy 3C\,33, whose poor environment is discussed in \citet{kra07b}.
If the lobes of 3C\,33 are expanding at a similar speed to those of
Cen A, and we assume (i) that the particle acceleration is similarly
efficient and (ii) that the density of accelerated particles is
proportional to the density of the shocked gas, then the expected
X-ray synchrotron flux should scale as $S \propto n_{ism} V
D_{L}^{-2}$. The external density ratio $n_{\rm 3C33}/n_{\rm CenA}$ is
$\sim 0.01$, the volume ratio is $\sim 7500$, and the square of the
ratio of luminosity distances is $1.9 \times 10^{-4}$, giving an
expected reduction in X-ray flux of a factor $\sim 70$ and a predicted
1-keV flux density from synchrotron emission of $\sim 1$ nJy. This is
comparable to the level of non-thermal X-ray emission measured by
Kraft et al. and attributed to inverse-Compton scattering of the
cosmic microwave background (CMB) \citep[e.g.][]{jhc05}, and so for
3C\,33 the observed X-ray flux levels from the radio lobes do not rule
out the possibility that high-energy particle acceleration is
occurring at a similar level to Cen A. However, the spectral index for
the non-thermal X-ray emission from 3C\,33 and other FRII radio lobes
($\sim 1.5 - 1.7$) is flatter than that of the shell in Cen A, and no
edge-brightening is observed for the X-ray lobes of FRIIs
\citep{jhc05}. In addition, the X-ray flux from this process decreases
as $(1 + z)^{4}$, unlike the IC/CMB flux, and so IC/CMB is predicted
to dominate by a large factor in the majority of FRIIs, which are more
distant than 3C\,33. Nevertheless, it is plausible that particle
acceleration by expanding FRII lobes may result in significant cosmic
ray populations and potentially detectable X-ray synchrotron emission
for sources in weaker environments.

\subsection{Implications for Cen A as a cosmic ray source}
\label{uhecr}
It is interesting to consider the implications of these results for
our understanding of Cen A as a source of cosmic rays, particularly in
light of recent results from the Pierre Auger Observatory showing an
overdensity of ultra-high energy cosmic rays (UHECRs) from the
direction of Cen A \citep{abr07,mos08}. To have efficient cosmic-ray
acceleration to high energies, the particle gyroradius cannot be
significantly larger than the size of the acceleration region. By the
criterion of \citet{hil84}, for an assumed acceleration region of
$\sim 300$ pc (the observed thickness of the shock front:
Section~\ref{dyn}) and $B = 7 \mu$G, typical particle energies cannot
exceed $\sim 2 \times 10^{18}$ eV. Hence magnetic fields significantly
higher than equipartition (e.g. $\sim 400 \mu$G) would be required for
the shock front to accelerate cosmic rays to $10^{20}$ eV. There is
considerable evidence that magnetic fields can be amplified by large
factors at the forward shocks of SNRs \citep[e.g.][]{bv04,ell04} (for
example lower limits on $B$ in Cass A are of the order of those
required here to confine CRs of $10^{20}$ eV: \citealt{vl03}), and so
if similar processes were operating at the Cen A shock front then
UHECR energies could be achieved. However, the equipartition magnetic
field strengths in SNRs are comparable to values inferred from the
thickness of the shock front. This suggests that magnetic field
amplification by similar factors at the Cen A shock may be unlikely,
as it would require a large departure from equipartition in the
direction of magnetic field dominance. The Cen A shock front appears
resolved in our {\it Chandra} observations, with a width of $\sim 2$
arcsec (corresponding to $\sim 35$ pc). We can use the shock thickness
to estimate the magnetic field strength in the shell, assuming it is
determined by synchrotron losses. Following the method used by
\citet{vl03} for the Cas A SNR, assuming the accelerated electrons
travel away from the shock front with the shocked gas at a speed of
$0.25v_{shock}$, we find synchrotron lifetimes of $\sim 5 \times
10^{4}$ y are required. These lifetimes for 1-keV photons imply
magnetic field strengths of $\sim 1 \mu$G, consistent with the
equipartition values dicussed above, but much lower than would be
required to accelerate UHECRs. In addition, as discussed above, our
X-ray observations suggest that we may be observing the high-energy
cut-off in the electron distribution directly, at energies of $\sim
\gamma = 10^{7} - 10^{8}$, which corresponds to electron energies of
$\sim 10^{14}$ eV. Nevertheless, if the cut-off energy is determined
by losses then its value will be higher for hadrons. We conclude that
it is possible that the Cen A shock could accelerate cosmic rays to
UHECR energies; however, this would require magnetic field
amplification by large factors, and magnetic domination of the
energetics of the synchrotron-emitting plasma. The $\sim
600$-kpc-scale giant outer lobes of Cen A may be a more likely
candidate for the origin of the UHECRs apparently associated with Cen
A \citep[e.g.][]{mos08,mjh08}.

\subsection{Implications for Cen A as a TeV source}

\begin{figure*}
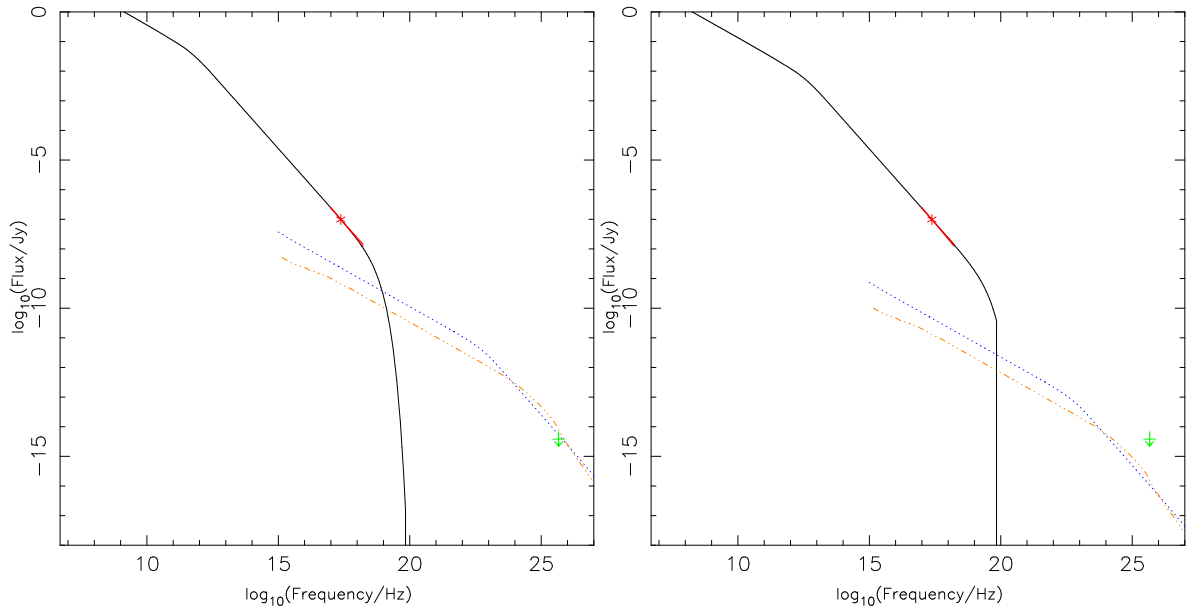

\centerline{\hbox{
\epsfig{figure=tev_1muG.ps,height=8cm}
\epsfig{figure=tev_7muG.ps,height=8cm}}}
\caption{Predictions for inverse-Compton emission from the SW shell of
  Cen A from scattering of starlight (orange dash-dotted line) and the
  CMB (blue dotted line). The electron distribution is as in the
  left-hand panel of Fig.~\ref{sedfig} for $B = 1\mu$G (left) and
  $7\mu$G (right). The star indicates the measured X-ray flux density,
  and the upper limit is the HESS measurement of \citet{aha05}.}
\label{tev}
\end{figure*}
 
It is also important to consider the implications of our results for
emission at higher energies. The detection of X-ray synchrotron
emission from the shell implies the existence of electrons with TeV
energies (for plausible values of $B$), and hence it is possible that
the shock could produce significant emission at very high energies. We
therefore investigated the predicted inverse-Compton emission from the
Cen A shell. The dominant photon fields in the shell region that can
be scattered to gamma-ray energies are (1) starlight from the host
galaxy, and (2) the cosmic microwave background (CMB). We determined
the photon density from starlight at the position of the shell by
deprojecting the V-band surface brightness profile of \citet{vdb76},
using the method of \citet{mm87} to obtain a V-band spectral energy
density at a radius of 268 arcsec (midway between the inner and outer
radii encompassing Region 1) of $1.6 \times 10^{-28}$ J Hz$^{-1}$
m$^{-3}$. A spectral energy distribution for the entire galaxy was
obtained using the UV measurements of \citet{wel79}, optical
measurements from \citet{dev91}, infrared measurements from 2MASS
\citep{jar03} and IRAS \citep{gol88}, which was normalised to the
V-band energy density given above. We note that the corresponding
total energy density at the radius of interest is in good agreement
with that determined by \citet{sta06}. The analysis was carried out
for Region 1, so as to include most of the X-ray-emitting electrons,
and the electron population was modelled as described in
Fig.~\ref{sed}. We used the {\sc synch} code of \citet{mjh98} to
determine the predicted inverse-Compton emission from starlight and
IC/CMB. Fig.~\ref{tev} shows the inverse-Compton predictions for
magnetic field strengths of 1 $\mu$G (the value obtained above from
consideration of the width of the X-ray shell) and $7\mu$G (the
equipartition value for Region 2 discussed in Section~\ref{sed}). Also
plotted is an upper limit from the High-Energy Stereoscopic System
(HESS) of $<5.68 \times 10^{-12}$ ph cm$^{-2}$ s$^{-1}$ above 0.19 TeV
\citep{aha05}, which corresponds to a flux density at 0.19 GeV of $3.8
\times 10^{-15}$ Jy for an assumed photon index of 2.0, consistent
with our assumed spectrum. For $B = 7\mu$G, we predict a total flux
density at 0.19 GeV from IC scattering of starlight and the CMB of
$3.2 \times 10^{-16}$ Jy, an order of magnitude below the HESS limit,
which corresponds to $F(>0.19 {\rm GeV}) \sim 2 \times 10^{-13}$
photons cm$^{-2}$ s$^{-1}$ above 0.19 TeV (for $\Gamma$ = 2). However,
with a lower magnetic field strength, as is suggested by the
calculation in Section~\ref{uhecr}, we find a predicted total flux
density at 0.19 GeV of $1.6 \times 10^{-14}$ Jy, corresponding to a
total $F(>0.19 {\rm GeV}) \sim 2 \times 10^{-11}$ ph cm$^{-2}$
s$^{-1}$, i.e. we predict a flux that is comparable to or higher than
the HESS upper limit. Hence it is reasonable to think that a TeV
detection of the Cen A shell may be possible with deeper observations.
It is unclear whether the Cen A shell would be expected to dominate
over emission from the jet (which also contain electrons of TeV
energies: e.g.\ \citealt{mjh07}) at TeV energies. A full calculation of
the predicted TeV emission from the jet is beyond the scope of this
paper; however, a rough estimate based on the X-ray jet properties
discussed in \citet{mjh07} suggests a level of TeV emission comparable
to that from the shell. In principle it would be possible to
distinguish spectrally between IC emission from the jet and the shell,
as the diffuse X-ray synchrotron emission in the jet has a
significantly steeper spectral index than the shell. Hence deeper TeV
observations of Cen A have the potential to provide important
constraints on the properties of the fluid in the shocked shell as
well as the jet.

\section{Conclusions}
\label{conclusions}

We have presented a detailed study of the X-ray emission associated
with the south-west inner lobe of Cen A. Our deep observations
as part of a {\it Chandra} Very Large Programme have enabled us to
establish firmly that the majority of the X-ray emission associated
with the shock surrounding the SW lobe, first discussed by Kraft et
al. (2003), is X-ray synchrotron emission associated with particle
acceleration at the shock front, and not thermal emission from shocked
gas. We have investigated the spectral structure of the X-ray shell,
presenting maps of spectral structure for the first time, and we
conclude that non-thermal emission dominates over most of the shell,
except in the inner regions, where the thermal X-ray emission can be
used to determine the shock conditions. We have described a
self-consistent model for the lobe and shock dynamics, considering in
detail the implications of possible electron-ion non-equilibrium in
the shocked material. Our analysis of the spatial distribution of
power-law spectral index constrains $\gamma_{max}$ for the accelerated
particles to be $\sim$10$^8$ at the strongest part of the shock,
consistent with the expectations of diffusive shock acceleration
theory. We considered the implications of our results for Cen A's role
as a source of cosmic rays, but conclude that the inner lobe shocks
are unlikely to be an important source of ultra-high energy cosmic
rays as this would require very high magnetic field strengths. Finally
we considered the expected inverse-Compton emission from this
synchrotron shell, finding that the predicted emission from IC
scattering of starlight and the CMB is comparable to current upper
limits at TeV energies.

\section*{Acknowledgments}

This work was partially supported by NASA grants NAS8-03060 and
GO7-8105X, and Hubble grant HST-GO-10597.03-A . We gratefully
acknowledge support from the Royal Society (research fellowship for
MJH).

\label{lastpage}
\end{document}